\documentclass[preprint,10pt]{elsarticle}

\usepackage{booktabs}
\usepackage{amsmath}
\usepackage{siunitx}
\usepackage{url}
\usepackage{hyperref}
\usepackage{multirow}
\usepackage{subcaption}

\journal{Medical Image Analysis}
\begin{document}
\begin{frontmatter}

\title{Domain adaptation strategies for cancer-independent detection of lymph node metastases}

\author{P\'eter~B\'andi}
\author{Maschenka~Balkenhol}
\author{Marcory~van~Dijk}
\author{Bram~van~Ginneken}
\author{Jeroen~van~der~Laak}
\author{and~Geert~Litjens}

\address{Nijmegen, the Netherlands}

\begin{abstract}
Recently, large, high-quality public datasets have led to the development of convolutional neural networks that can detect lymph node metastases of breast cancer at the level of expert pathologists. Many cancers, regardless of the site of origin, can metastasize to lymph nodes. However, collecting and annotating high-volume, high-quality datasets for every cancer type is challenging. In this paper we investigate how to leverage existing high-quality datasets most efficiently in multi-task settings for closely related tasks. Specifically, we will explore different training and domain adaptation strategies, including prevention of catastrophic forgetting, for colon and head-and-neck cancer metastasis detection in lymph nodes.

Our results show state-of-the-art performance on both cancer metastasis detection tasks. Furthermore, we show the effectiveness of repeated adaptation of networks from one cancer type to another to obtain multi-task metastasis detection networks. Last, we show that leveraging existing high-quality datasets can significantly boost performance on new target tasks and that catastrophic forgetting can be effectively mitigated using regularization.
\end{abstract}

\begin{keyword}
cancer, lymph node, deep learning, convolutional neural network, domain adaptation
\end{keyword}

\end{frontmatter}


\section{Introduction}
Cancer can originate from different sites and subsequently have many different appearances, ranging from skin melanomas to prostate adenocarcinomas. Many cancer types have the tendency to metastasize to the lymphatic system. The spread of cancer from its initial site is a hallmark of progressive disease and associated with a poor prognosis. As such, the presence of lymph node metastases is an important part of the internationally accepted classification system for solid tumours: the TNM staging system \citep{Sobin11}. Lymph node metastases for many primary cancer sites, such as breast, colon, and head-and-neck cancer, are diagnosed through histopathological inspection of biopsied or surgically removed lymph nodes of cancer patients. Generally, this is a tedious and time-consuming task for pathologists, and sometimes (mostly smaller) metastases are overlooked.

In recent years, thanks to the advent of whole-slide scanners, deep convolutional neural networks (CNNs) and availability of large public datasets \citep{Litjens18}, algorithms have been developed that can detect lymph node metastases for breast cancer at the level of expert pathologists \citep{Ehteshami17}. Such systems can aid pathologists in identifying metastases more quickly and more accurately \citep{Liu18}. However, training of accurate, reliable CNNs generally requires large, high-quality datasets. Collecting such well-curated datasets for every single cancer type would be an extremely difficult task.

Luckily, many of these detection tasks are closely related. As an example, normal lymph node tissue is similar throughout the human body, so the `non-metastasis' class is the same for every cancer type. As such, we hypothesize that leveraging existing high-quality datasets for a specific cancer type might help train accurate CNNs for another type. Furthermore, tumours from different sites that originate from the same cell type have a similar appearance. For example, most breast and prostate cancer are adenocarcinomas, meaning they originate from the epithelial cells of glandular tissue.

The most basic strategy to handle multiple tasks is to simply combine two datasets and train a single deep convolutional network to accomplish both tasks in one go. However, this strategy has several shortcomings, first and foremost that sufficient training data for both tasks needs to be available. Secondly, retraining a deep learning system from scratch is typically a computationally inefficient solution. Last, there is no guarantee that a system trained on both tasks simultaneously will be optimal for either. In literature, several more sophisticated strategies are presented, generally referred to under the moniker `transfer learning'. When, as in our case, the tasks are closely related, the term `domain adaptation' is typically used.

Roughly, the basic concept is that we start from the weights of the network trained on a `source task' and then continue training (i.e.\ fine-tune) on the training data of the `target task'. This tends to work very well, as has been shown in numerous publications using pre-trained networks on the ImageNet dataset for a variety of tasks \citep{Campanella19_08, Bandi18}. A key challenge in transfer learning is that networks, while fine-tuning on the `target task' tend to show severely deteriorated performance on the `source task', so-called catastrophic forgetting. Identifying strategies to prevent this phenomenon is one of the goals of the field of `continual learning'. We give an overview of existing `transfer learning' and `continual learning' strategies in the related work section.

In this paper we will investigate various forms of `domain adaptation' in the context of detection of lymph node metastases for varying cancer types in digitized histological sections. Specifically, we will use the large, open-access CAMELYON lymph node dataset composed of 1399 slides \citep{Ehteshami17, Bandi18, Litjens18} for the `source task' of breast cancer metastasis detection. Additionally, we collected two more lymph node datasets for colon cancer and head-and-neck cancer metastases detection as `target tasks'. We also investigate repeated `domain adaptation' to allow not just dual- but also multi-task networks.

Our main contributions are the following
\begin{itemize}
    \item We present state-of-the-art performance on colon and head-and-neck cancer metastasis detection.
    \item We show the effectiveness of repeated `domain adaptation' to obtain multi-task metastasis detection networks.
    \item We show that leveraging existing high-quality datasets in a transfer learning setting can significantly boost performance on new `target tasks' in the context of cancer metastasis detection.
    \item We show that catastrophic forgetting is also an issue for closely related `source' and `target tasks' and can be effectively mitigated with elastic weight consolidation.
\end{itemize}

\section{Related Work}
\textit{Transfer learning} is a machine learning technique that focuses on storing knowledge gained while solving a problem and applying it to a different but related problem \citep{Thrun12}. It is a popular approach in deep learning where pre-trained models on a certain task, e.g.\ natural image classification, are used as the starting point for training on a different task, e.g.\ prostate cancer detection in histopathology \citep{Campanella19_08}. It has been shown that fine-tuning a model that was pre-trained on high-quality datasets (e.g.\ ImageNet) is a simple way to improve performance on a `target task' for which training data is scarce or labels are noisy \citep{Jiang20}.

\textit{Continual learning} is the ability of artificial intelligence systems to learn consecutive tasks without forgetting how to perform previously learned tasks. Continual learning is particularly challenging for artificial neural networks since they have the tendency to lose knowledge of previously learned tasks while the information that is relevant to their `target task' is incorporated. This phenomenon, called \textit{catastrophic forgetting} occurs when a CNN is trained sequentially on multiple tasks, where the weights in the network that are important for the first task are adapted to meet the objectives of the subsequent tasks \citep{McCloskey89, French99}.

The different methods of preventing catastrophic forgetting can be classified into three categories: dynamic architectures, rehearsal strategies, and regularization approaches \citep{Chen18, Parisi19}. 

\textit{Dynamic architectures} extend the network in some way for each new task. \textit{Progressive Neural Network} is a dynamic architecture that blocks any changes to the network trained on previous knowledge and expand the architecture by allocating subnetworks with fixed capacity to be trained with the new information. It retains a pool of pre-trained models, one for each learned task \citep{Rusu16}. Instead of extending the architecture \textit{Expert Gate} proposes a network of experts where each expert is a model trained given a specific task and a set of gating autoencoders that learn a representation for the task at hand, and, at test time, automatically forward the test sample to the relevant expert \citep{Aljundi17}. Another dynamic approach is \textit{Dynamically Expandable Networks} (DEN) that increases the number of trainable parameters to incrementally learn new tasks. DEN is trained in an online manner by performing selective retraining which expands the network capacity using group sparse regularization to decide how many neurons to add at each layer \citep{Lee17}. A downside of dynamic architectures is that typically the network (part) responsible for the `source task' is frozen, preventing it from improving given data from a related `target task' such as in `domain adaptation'.

\textit{Rehearsal strategies} try to maintain or generate examples of previously learned tasks and interleave them with the examples of the new tasks during training. The \textit{Incremental Classifier and Representation Learning} method maintains a set of exemplar examples for each observed class. For each class, an exemplar set is a subset of all examples of the class, aiming to carry the most representative information of the class \citep{Rebuffi17}. The \textit{Deep Generative Replay} approach uses dual-model architecture consisting of a deep generative model and a task solver. It samples training data from previously learned tasks in terms of generated pseudo-data and interleaves with information from the new task \citep{Hanul17}. The disadvantage of the former approach is that (part of) the original dataset needs to be available in addition to the pre-trained weights of the network. For the latter approach a challenge in the context of lymph node metastases is that changes between different cancer classes are relatively subtle, and as such might be difficult to capture with a generative adversarial network.

\textit{Regularization approaches} try to limit the change of the parameters that are most relevant for previous tasks, and use the ones that are less relevant to learn the `target task'. The \textit{Elastic Weight Consolidation} (EWC) method consists of a quadratic penalty on the difference between the parameters for the old and the new tasks that slows down the learning for task-relevant weights coding for previously learned knowledge. The relevance of the parameters for a previous task is approximated as a Gaussian distribution with mean given by their learnt value for the task and a diagonal precision given by the diagonal of the Fisher information matrix. The Fisher information matrix is calculated for each individual task \citep{Kirkpatrick17}. The \textit{Synaptic Intelligence} (SI) approach alleviates catastrophic forgetting by allowing individual synapses (i.e.\ neurons) to estimate their importance for solving a learned task. Similarly to EWC the approach penalizes changes to the most relevant synapses so that new tasks can be learned with minimal forgetting \citep{Zenke17}. A disadvantage is the need to distribute some extra parameters per weight in addition to their value, but in terms of data size this is much less than providing (a subset of) the training data.

In this work we used EWC to prevent catastrophic forgetting and compared it to training strategies on combined or gradually extended datasets.

\section{Materials}

\subsection{Collection of whole-slide images}
In total we collected 1667 whole-slide images (WSIs) of hematoxylin and eosin (H\&E) stained glass slides of lymph node tissue. Of the 1667 WSIs, 1399 were from the CAMELYON dataset for breast cancer metastasis detection \citep{Litjens18}, 149 slides were collected for colon cancer metastasis detection and 119 slides were collected for head-and-neck cancer metastasis detection. These datasets will be referred to as the breast cancer, colon cancer, and head-and-neck cancer dataset, respectively. Additional details for each of these subsets are presented in their respective subsections.

All WSIs contained multiple resolution levels, with approximately \num{1e5} by \num{2e5} pixels at the highest resolution level. Each consecutive resolution level doubled the pixel size in both directions and halved the pixel count in each dimension. The typical file size of a WSI was 3.5 GB, but it varied greatly depending on the tissue content of the image. Pixel sizes differed slightly depending on the scanner used for digitization, but are all within \SIrange{0.24}{0.25}{\um}. Details for all subsets are also presented in Table \ref{table:datasets}.

\subsubsection{Breast cancer dataset}
Clinically, three categories of metastases are distinguished in sentinel lymph nodes of breast cancer, based on their size: macro-metastases, micro-metastases and isolated tumor cells (ITCs). When multiple metastases are present, the metastasis with the largest size determines the metastasis category of the slide. For each slide, either exhaustive annotations of the metastases were given (CAMELYON16, subset of CAMELYON17) or a slide-level label of the largest metastases was provided (most of CAMELYON17).

For algorithm training and validation, we split the breast cancer dataset into a CNN development set, a random forest classifier (RFC) development set, and a testing set \citep{Breiman01}. We used the 270 WSIs from the original training set of the CAMELYON16 challenge plus 50 randomly selected negative slides and the 50 positive slides with exhaustively annotated metastases from the original training set of the CAMELYON17 challenge as the CNN development set. These 370 images were divided into training and validation subsets with 291 and 79 WSIs, respectively.

The remaining 400 slides from the original training set of the CAMELYON17 challenge that were not part of the CNN development set were used for training and validating RFCs for post-processing CNN output. The RFCs were used to classify WSIs into one of the lesion classes of the CAMELYON17 challenge. The RFCs used features extracted from the metastasis likelihood maps obtained after CNN inference. The 400 WSIs were divided into RFC training, and RFC validation subsets with 280 and 120 slides respectively. The training part was composed of 186 negative and 94 positive WSIs, while the validation part had 82 negative and 38 positive WSIs. For both the CNN and RFC development sets, the images were divided between the two subsets randomly.

The 129 WSIs from original test set of the CAMELYON16 challenge and the 500 test images from the CAMELYON17 challenge were exclusively used for algorithm evaluation. We refer to these test sets as \textit{CAMELYON16 test set} and \textit{CAMELYON17 test set} respectively.

When randomly assigning slides to each subset, we used the available metastasis labels to keep the ratio across subsets similar. Furthermore, we stratified according to medical center to ensure accurate representation of all five centers present in the CAMELYON dataset in both sets. For extensive details on the CAMELYON dataset, we refer to \citep{Litjens18}.

\subsubsection{Colon cancer dataset}
In colon cancer diagnostics, no distinction is made between metastasis sizes. As such, for each slide a binary label was acquired: metastasis or no metastasis. Additionally, detailed annotations of all metastases were made, see subsection \ref{sec:annotations} for more details.

The colon dataset consisted of 149 WSIs of lymph nodes of colon cancer patients from Rijnstate Hospital in Arnhem, the Netherlands. The dataset was divided into training, validation and testing subsets with 75, 30, and 44 WSIs, respectively. The WSIs were assigned to the subsets randomly while balancing for the presence of metastases. All slides were digitized with a 3DHistech Pannoramic Flash II 250 at a resolution of \SI{0.24}{\um}.

\subsubsection{Head-and-neck cancer dataset}
For the head-and-neck slides, a slide-level binary label was available in addition to detailed lesion annotations, similar to the colon dataset. 

The head-and-neck cancer dataset consisted of the 119 WSIs of lymph nodes of head-and-neck squamous cell carcinoma patients from two different medical centers, Amsterdam University Medical Center and the Radboud University Medical Center. The dataset was divided into training, validation and testing subsets with 60, 24, and 35 WSIs, respectively. As with the other datasets, the distribution of the images was balanced according to center and presence of metastases. The slides were digitized with a 3DHistech Pannoramic Flash II 250 at a resolution of \SI{0.24}{\um}.

\begin{table}
\centering
\begin{tabular}{lrrrrrrr}
\toprule
Dataset       & Total & \multicolumn{2}{c}{Training} & \multicolumn{2}{c}{Validation} & \multicolumn{2}{c}{Testing} \\
              & & Neg & Pos & Neg & Pos & Neg & Pos \\
\midrule
Breast from CAMELYON16 & 399 & 127 &  89 &  32 &  22 &  80 &  49 \\
Breast from CAMELYON17 & 600 &  40 &  35 &  10 &  15 & 260 & 240 \\
Colon                  & 149 &  41 &  34 &  17 &  13 &  25 &  19 \\
Head-and-neck          & 119 &  37 &  23 &  15 &   9 &  22 &  13 \\
\bottomrule    
\end{tabular}
\caption{Composition of the datasets}
\label{table:datasets}
\end{table}

\subsection{Detailed lesions annotations}\label{sec:annotations}
\begin{figure}
\centering
  \begin{subfigure}{0.33\textwidth}
    \includegraphics[width=0.96\textwidth]{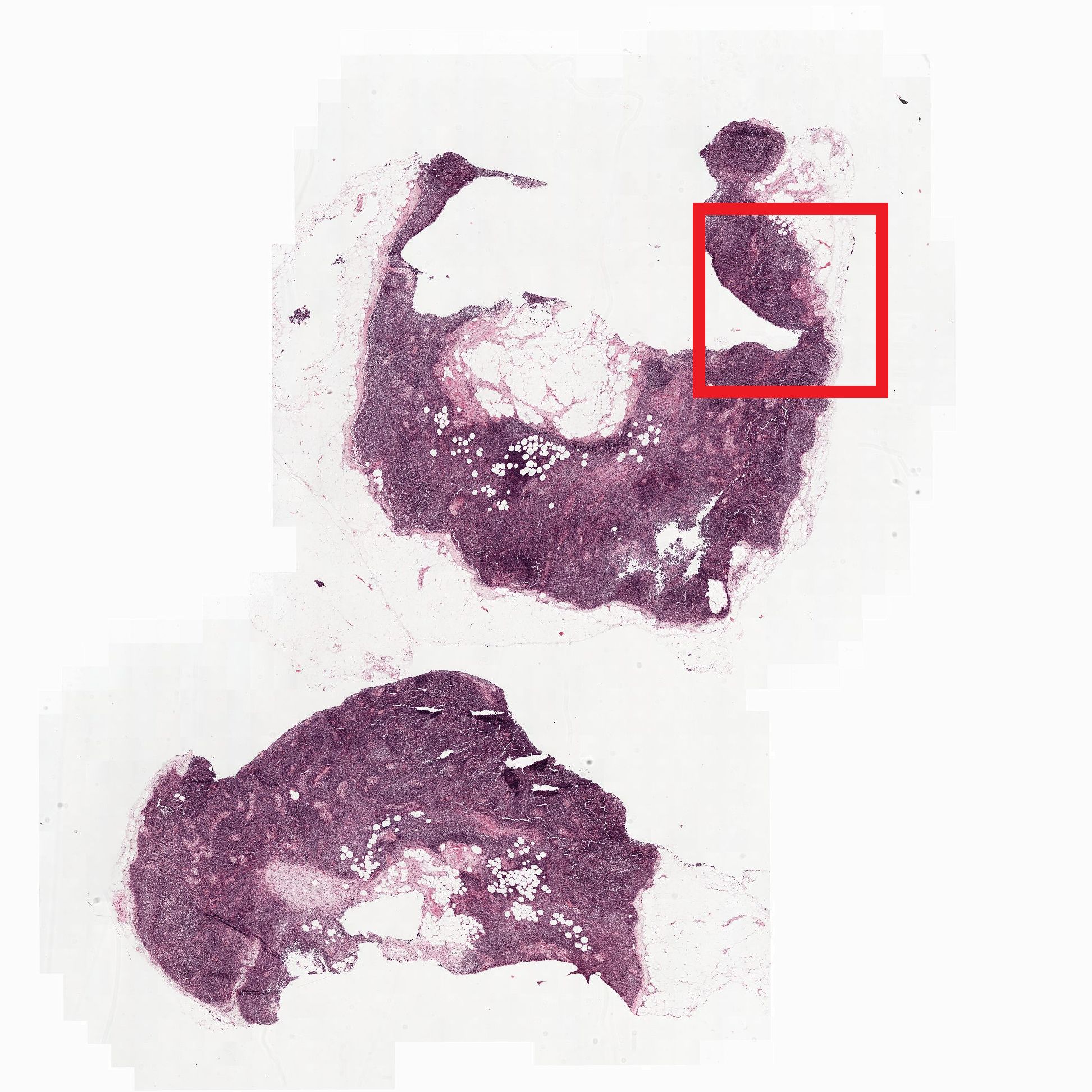}
    \vspace{3.6pt}
  \end{subfigure}%
  \begin{subfigure}{0.33\textwidth}
    \includegraphics[width=0.96\textwidth]{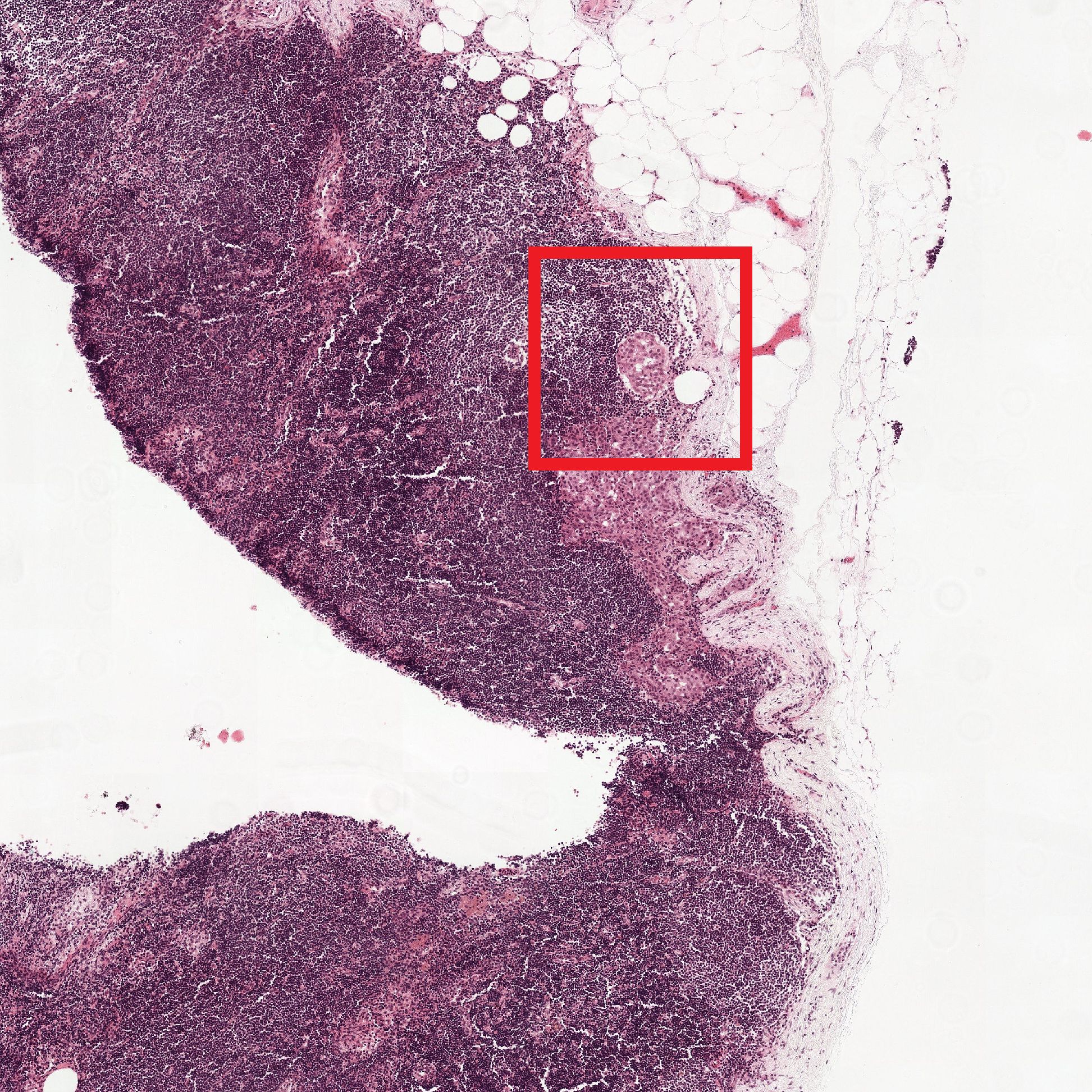}
    \vspace{3.6pt}
  \end{subfigure}%
  \begin{subfigure}{0.33\textwidth}
    \includegraphics[width=0.96\textwidth]{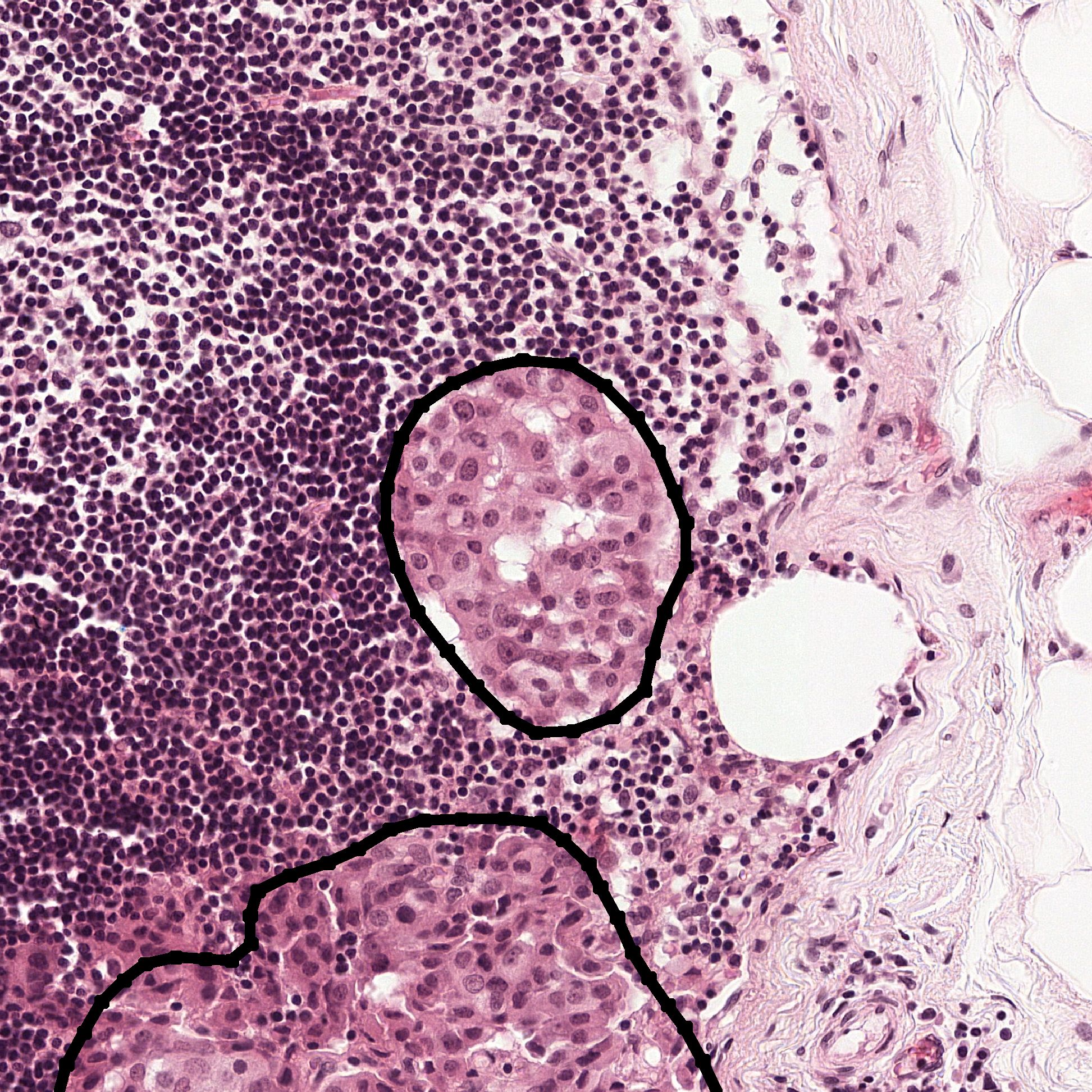}
    \vspace{3.6pt}
  \end{subfigure}\\
  
  \begin{subfigure}{0.33\textwidth}
    \includegraphics[width=0.96\textwidth]{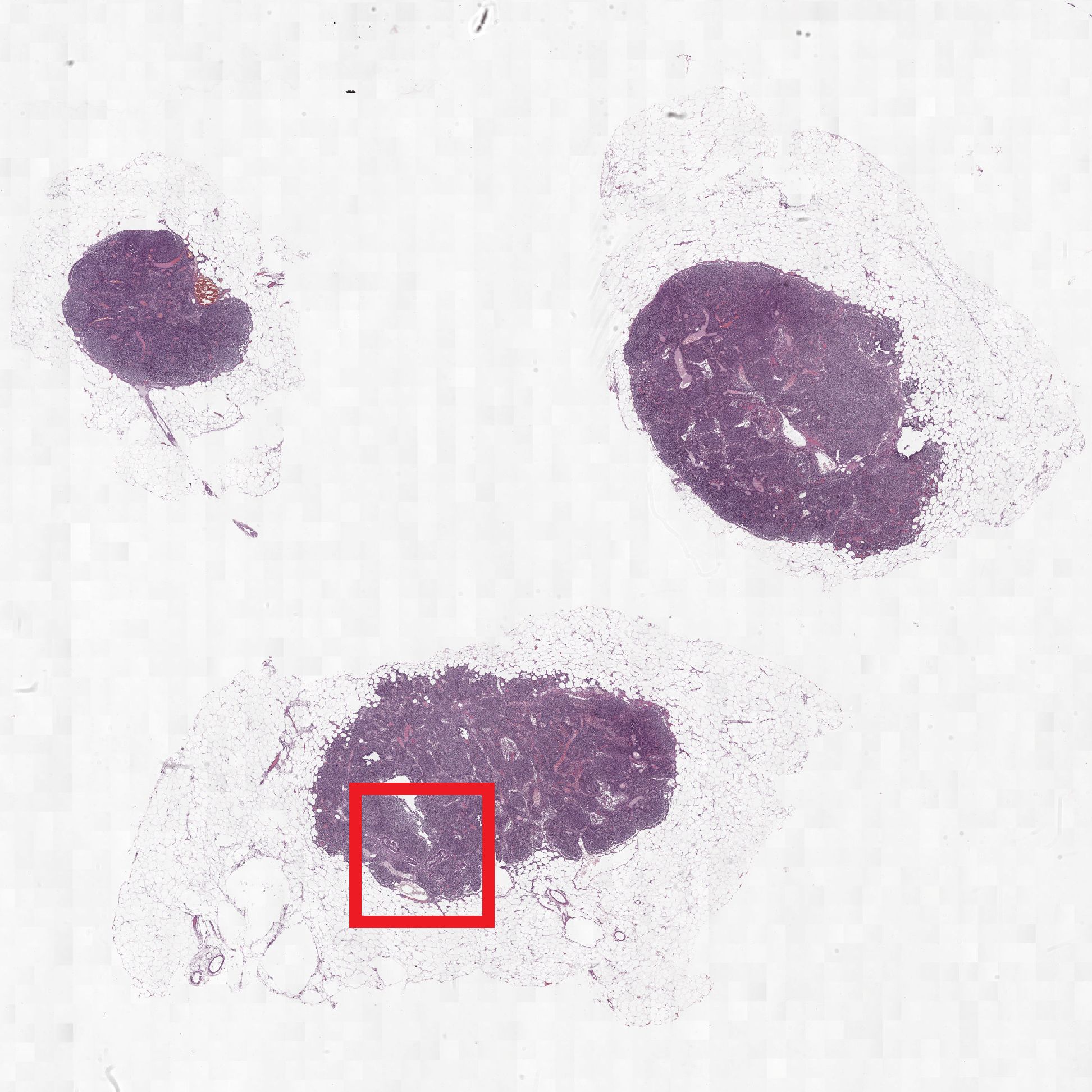}
    \vspace{3.6pt}
  \end{subfigure}%
  \begin{subfigure}{0.33\textwidth}
    \includegraphics[width=0.96\textwidth]{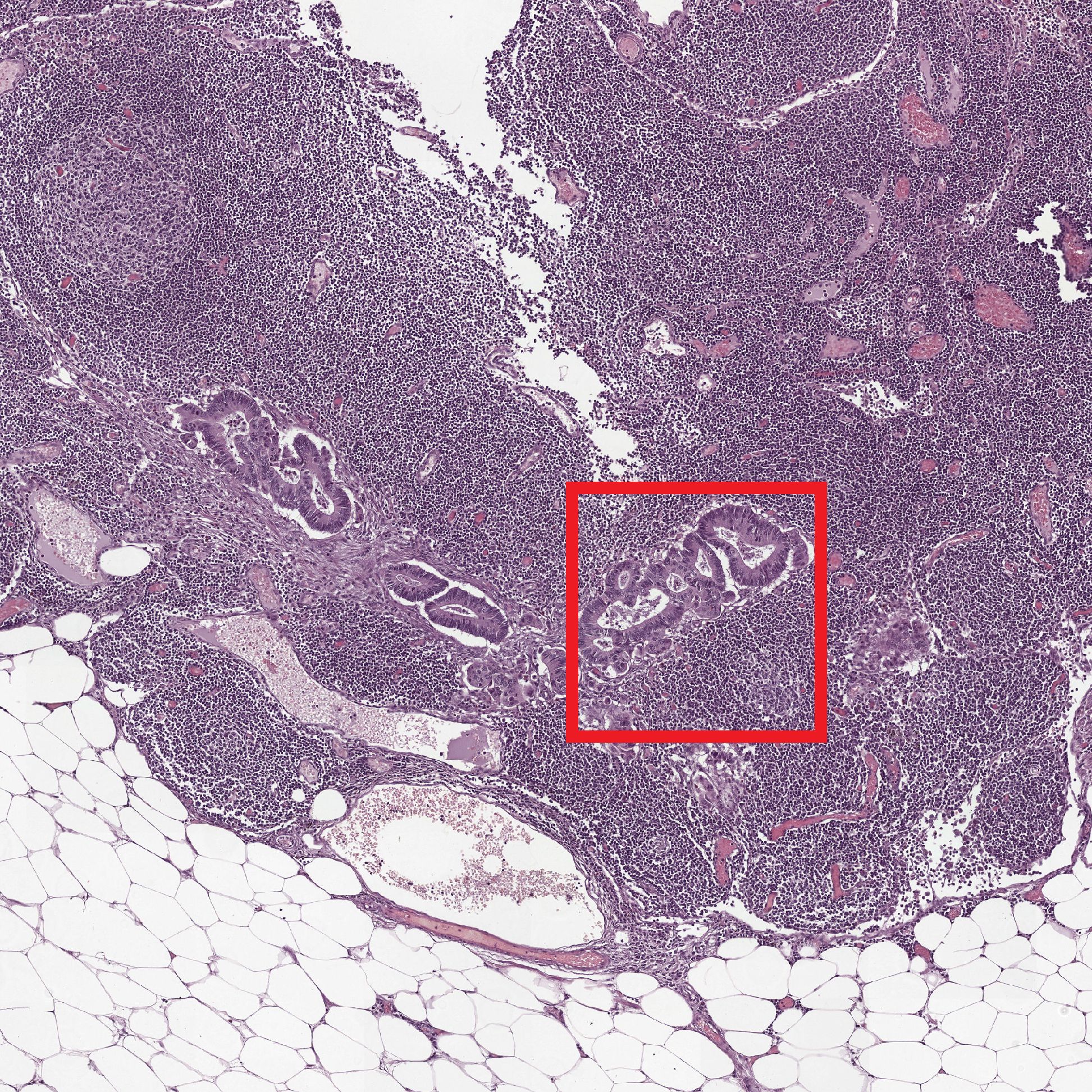}
    \vspace{3.6pt}
  \end{subfigure}%
  \begin{subfigure}{0.33\textwidth}
    \includegraphics[width=0.96\textwidth]{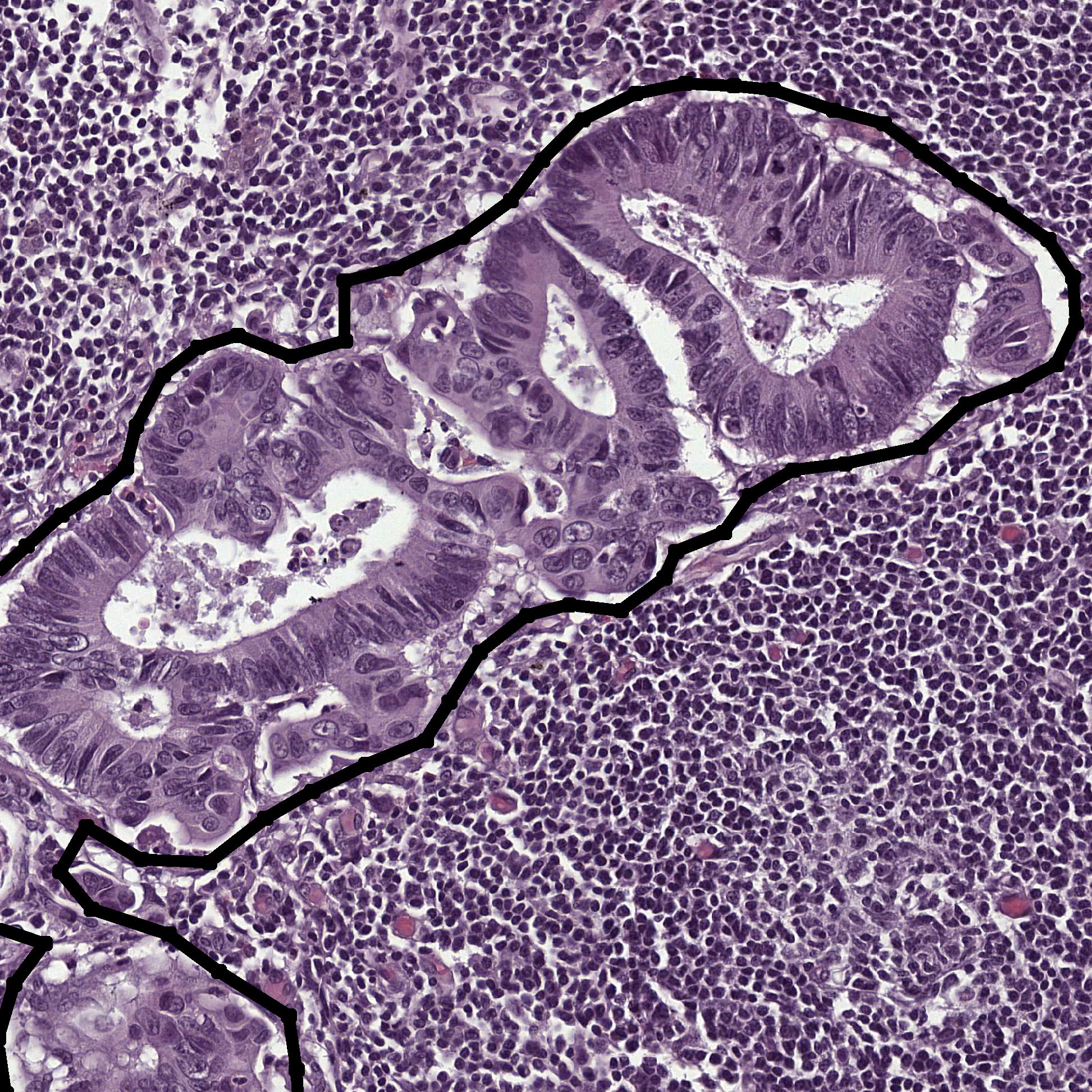}
    \vspace{3.6pt}
  \end{subfigure}\\
  
  \begin{subfigure}{0.33\textwidth}
    \includegraphics[width=0.96\textwidth]{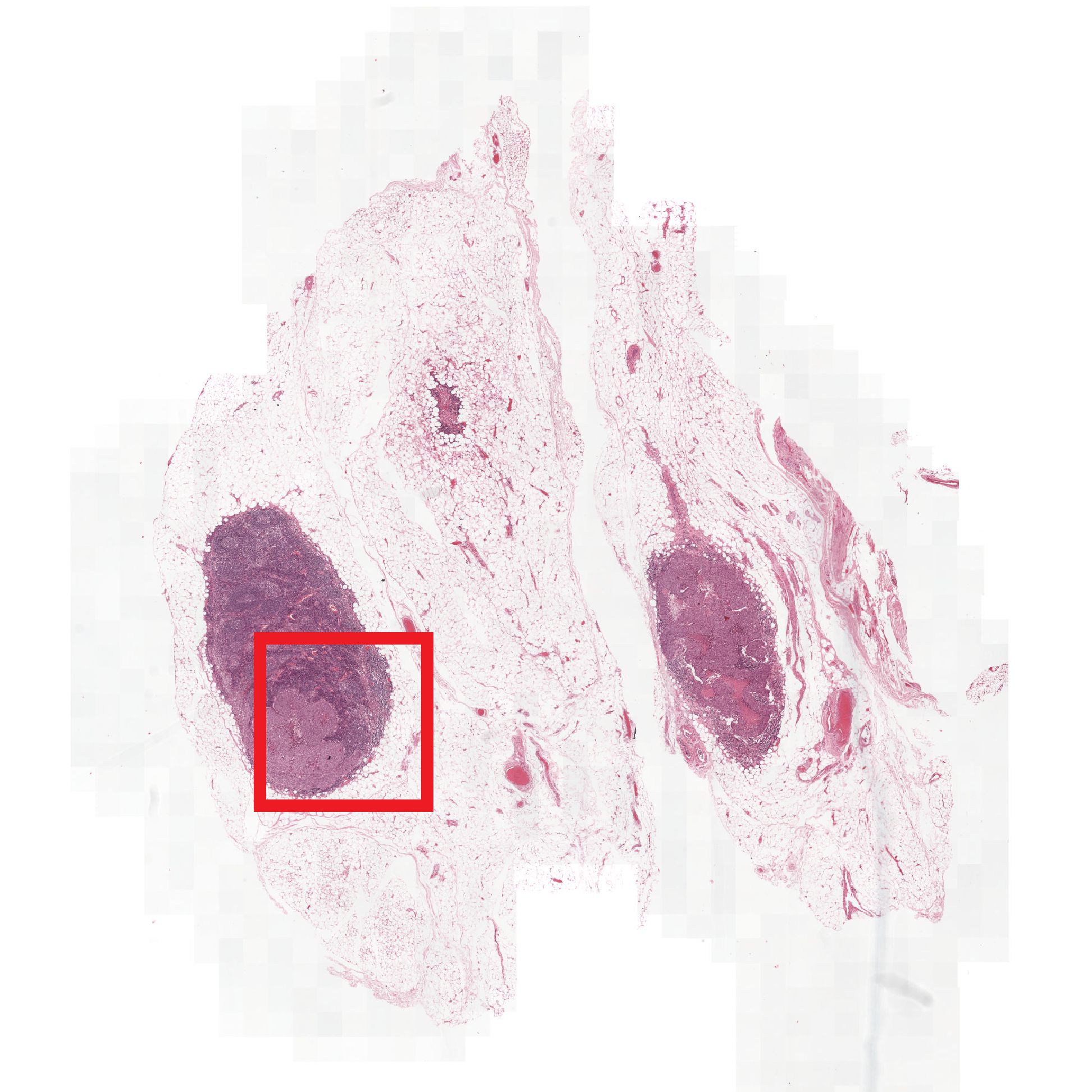}
  \end{subfigure}%
  \begin{subfigure}{0.33\textwidth}
    \includegraphics[width=0.96\textwidth]{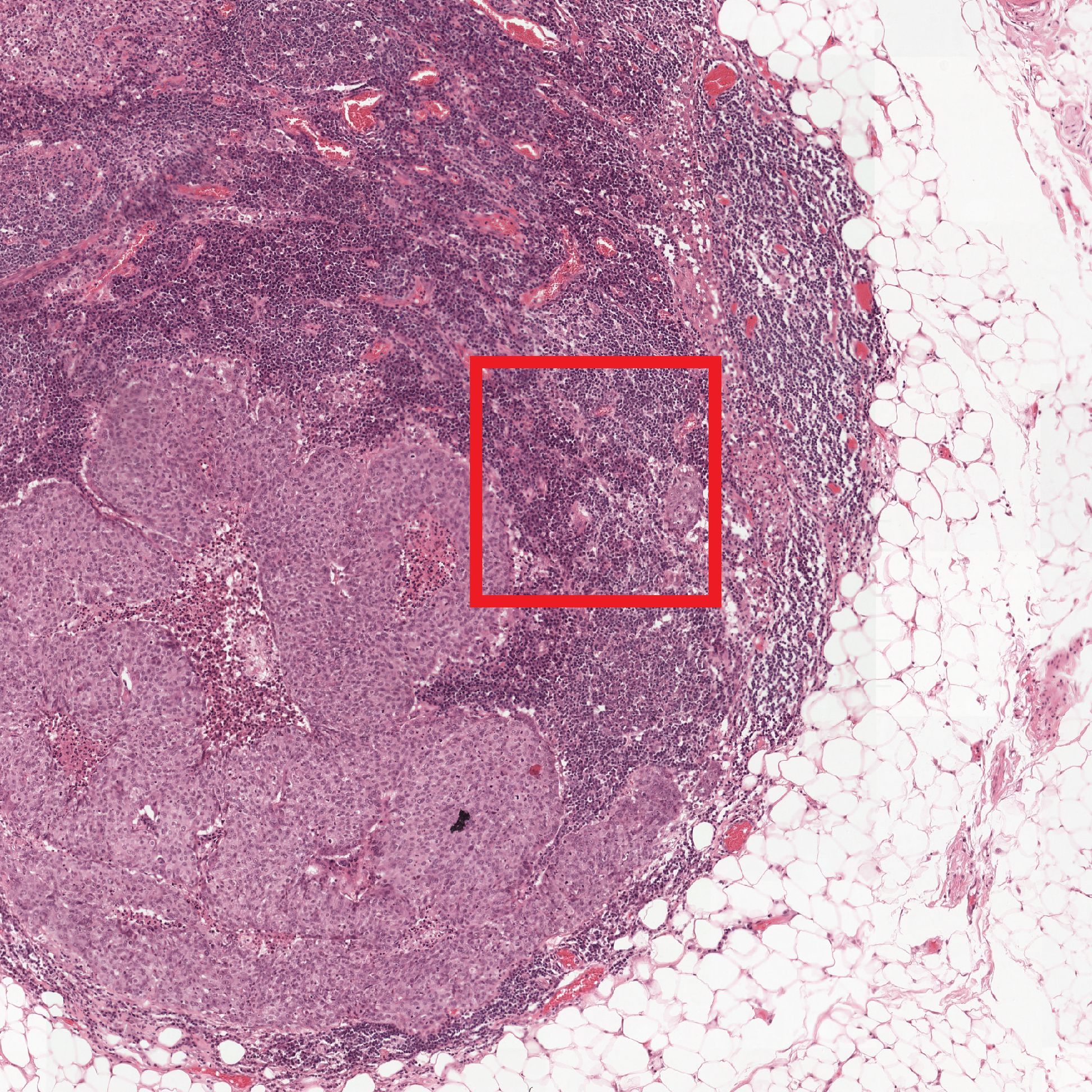}
  \end{subfigure}%
  \begin{subfigure}{0.33\textwidth}
    \includegraphics[width=0.96\textwidth]{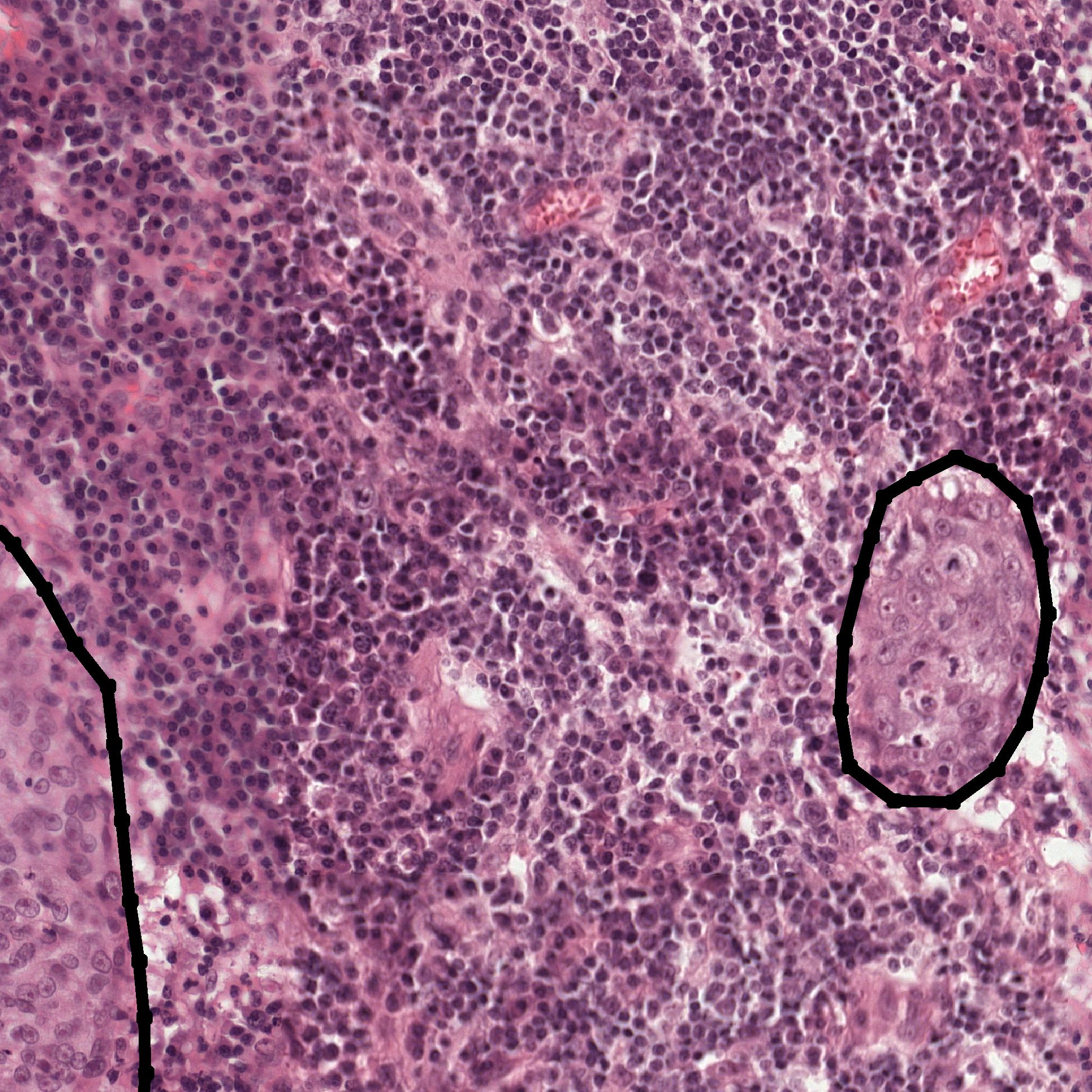}
  \end{subfigure}\\
  \captionsetup{justification=centering}
  \caption{WSI Examples in with increasing zoom levels (from left to right). Rows in order: lymph node sections of patients with breast, colon, head \& neck cancer respectively. The metastases are delineated at the highest zoom levels with black outline.}
  \label{figure:tissue_examples}
\end{figure}

From the CAMELYON16 and CAMELYON17 datasets we used the provided detailed annotations for both the CAMELYON16 and CAMELYON17 subsets \citep{Litjens18}. For the colon and head-and-neck cancer datasets individual metastatic lesions were outlined in detail using the ASAP digital pathology viewer software (version 1.9, available as open-source software on \href{https://github.com/computationalpathologygroup/ASAP}{GitHub}). The colon dataset was annotated by a pathologist (M.v.D.) while the head-and-neck dataset was annotated under supervision of a pathology resident (M.B.). Examples of the resultant annotations are shown in Figure \ref{figure:tissue_examples}. As part of the standard diagnostic practice in the Netherlands immunohistochemical (IHC) staining for cytokeratin can be used for clarification of unclear diagnosis on H\&E \citep{Chag05, Reed09}. When available, these corresponding cytokeratin-stained slides were used as an additional reference for annotating the lesions. 

\section{Methods}

In order to compare the different strategies we trained CNNs to predict if an image patch sampled from a WSI contained cancer. We used shift-and-stitch method to build complete probability maps for each WSI. Using these probability maps we calculated the same measurements that were used in the CAMELYON16 and CAMELYON17 challenges. We used RFCs to predict slide level classes from features extracted from the WSI probability maps then combined them to the case level classes of CAMELYON17. The following section details the network architecture, the training methods, and the used metrics.

\subsection{Networks Architecture}
For all experiments we used the same network architecture: a version of the DenseNet \citep{Huang17} with two modifications. We chose DenseNet as the base architecture as it currently holds the state-of-the-art result in the CAMELYON17 challenge \citep{Bandi18}.

First, we replaced the zero-padding in the DenseNet architecture with valid-padding so the network can be used efficiently for inference in a fully-convolutional manner \citep{Shelhamer17}. Using a network with zero-padding in a fully-convolutional manner results in distribution shift of feature values due to reduced contribution of the padded zeros, making prediction unreliable.

Input sizes of \begin{math}299 \times 299\end{math} pixels has been shown to work well with histopathological images at a resolution of \SI{0.5}{\um} \citep{Liu18}. However, when replacing zero-padding with valid-padding the input size needed to obtain the same output size grows exponentially. Because of this, we had to limit the depth of the network to retain workable input size. Through a series of preliminary experiments on the breast cancer development set we selected a DenseNet architecture with input size of \begin{math}279 \times 279\end{math} pixels, composed of 3 dense blocks, each with 4 \begin{math}1 \times 1\end{math} convolutions, followed by \begin{math}3 \times 3\end{math} convolutions. To correct for the size differences due to the lack of zero padding, cropping layers were added to the dense blocks to crop the in-block skip connections to the same size, so they could be concatenated. The transition blocks were composed of a \begin{math}1 \times 1\end{math} convolution layer followed by \begin{math}2 \times 2\end{math} average pooling.

In the dense blocks each \begin{math}1 \times 1\end{math} convolutional layer had 64 filters and each \begin{math}3 \times 3\end{math} convolutional layer had 32 filters. In the transition blocks the compression ratio was \begin{math} 0.5 \end{math} halving the number of filters coming from the preceding dense block.

\begin{figure}
  \centering
  \includegraphics[width=1.0\textwidth]{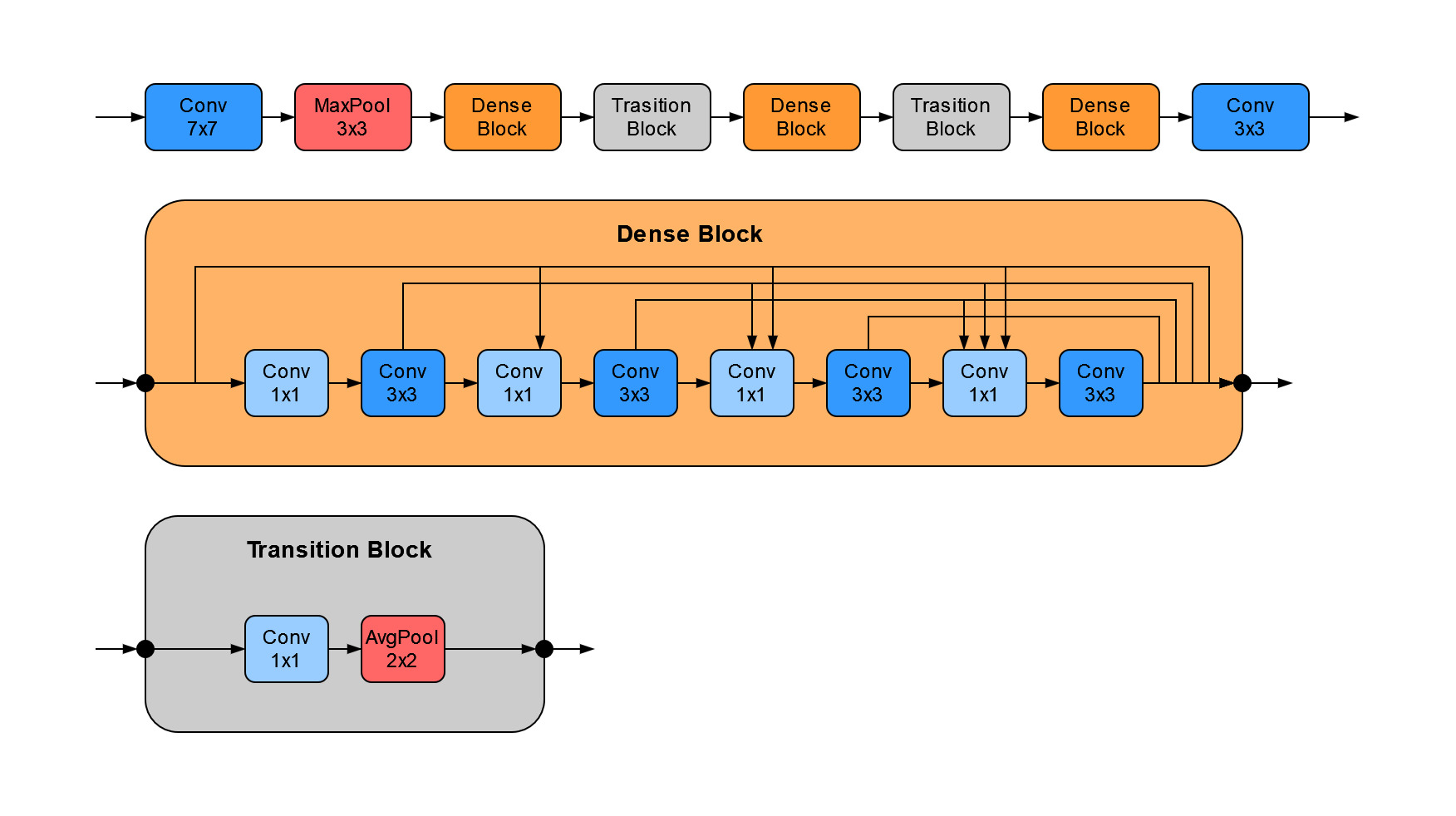}
  \caption{Network architecture.A version of the DenseNet model with an initial \begin{math}7 \times 7\end{math} convolution and a \begin{math}3 \times 3\end{math} max pooling followed by 3 dense blocks with transition blocks between them closed by a final \begin{math}3 \times 3\end{math} convolutional layer. Each dense block was composed of 4 \begin{math}1 \times 1\end{math} convolutions, followed by \begin{math}3 \times 3\end{math} convolutions. The transition blocks were composed of a \begin{math}1 \times 1\end{math} convolution layer followed by \begin{math}2 \times 2\end{math} average pooling.}
  \label{figure:network_diagram}
\end{figure}

We used batch normalization \citep{Ioffe15} and the ReLU \citep{Maas13} non-linearity after each convolutional layers except the last where we did not used batch normalization and the non-linearity was a soft-max. The network architecture is summarized in Fig. \ref{figure:network_diagram}.

\subsection{Network Training}
The networks were initialized with the He \citep{He15} initialization method and the weights were updated using the Adam optimizer \citep{Kingma14}. We used categorical cross-entropy as the loss function and added L2 regularization with a weight of \begin{math} \lambda = 10^{-4} \end{math}. Classification accuracy on the validation set was used to evaluate the network performance. The \begin{math} l = 10^{-4} \end{math} initial learning rate was divided by 10 after each four consecutive epochs without improvement, and the training procedures were stopped after 20 consecutive epochs without improvement. Both the training and the validation epochs were composed of 262,144 patches sampled from randomly selected positions from the WSIs, with a batch size of 32. 

For training and validation we extracted patches of \begin{math} 279 \times 279 \end{math} pixels selected randomly from the WSIs at a resolution of approximately \SI{0.5}{\um}. For each patch we assigned the label of the central pixel, either metastasis or normal tissue. The WSIs were pre-processed with a tissue-background segmentation CNN, so we could omit the non-tissue or empty areas from being selected \citep{Bandi19}. The patches were sampled with \begin{math}4:1\end{math} normal to cancer label ratio \citep{Liu18}. For experiments where the training and validation sets were a combination of multiple datasets (e.g.\ breast and colon) sampling was performed such that each dataset contributed equally to the training process (see the Experiments subsection for more details).

We used data augmentation during training to make our networks more robust to data variations. The loaded image patches were subjected to a series of augmentation steps with randomized parameters from pre-defined ranges with uniform distribution for each individual patch. Our augmentation methods greatly overlap with the published \textit{HSV-light} method \citep{Tellez19}. However, we did not use elastic transformation, the individual augmentation methods are executed in different order in our pipeline, and the ranges of the parameters were extended for greater variability. The augmentation pipeline consisted of horizontal mirroring, 90 degree rotations, scaling, color adjustment in hue-saturation-brightness color space, contrast adjustment, additive Gaussian noise, and Gaussian blur. Table \ref{table:augmentation_parameters} lists the augmentation steps and their parameter ranges.

\begin{table}
  \centering
  \begin{tabular}{cccc}
  \toprule
  \textbf{Order} &\textbf{Augmentation}    & \textbf{Parameter}               & \textbf{Range}                                                 \\
  \midrule
   1             & mirroring               & \begin{math} f \end{math}        & horizontal, none                                               \\
   2             & rotation                & \begin{math} \alpha \end{math}   & \begin{math}0^\circ, 90^\circ, 180^\circ, 270^\circ \end{math} \\
   3             & scaling                 & \begin{math} z \end{math}        & \begin{math} [0.9, 1.1] \end{math}                             \\
   4             & hue adjustment          & \begin{math} h \end{math}        & \begin{math} [-0.1, 0.1] \end{math}                            \\
   5             & saturation adjustment   & \begin{math} s \end{math}        & \begin{math} [-0.25, 0.25] \end{math}                          \\
   6             & brightness adjustment   & \begin{math} b \end{math}        & \begin{math} [-0.25, 0.25] \end{math}                          \\
   7             & contrast adjustment     & \begin{math} c \end{math}        & \begin{math} [-0.25, 0.25] \end{math}                          \\
   8             & Additive Gaussian noise & \begin{math} \sigma_a \end{math} & \begin{math} [0.0, 0.05] \end{math}                            \\
   9             & Gaussian blur           & \begin{math} \sigma_b \end{math} & \begin{math} [0.0, 1.0] \end{math}                             \\
  \bottomrule
  \end{tabular} 
  \caption{Augmentation configuration}  
  \label{table:augmentation_parameters}
\end{table}

\subsection{Elastic Weigh Consolidation}
For experiments in which we used elastic weight consolidation, we calculated the Fisher information matrix on the training sets \citep{Huszar17, Kirkpatrick17}. The relevance of the weights for a given task is estimated by the diagonal of the calculated Fisher information matrix. The \begin{math}\mathcal{L}_{EWC}(\theta)\end{math} EWC loss is calculated according to the following formula:

\begin{equation}
\mathcal{L}_{EWC}(\theta) = \sum_{i} \phi F_{i} (\theta_{i} - \theta_{i}^{*})^{2}
\end{equation}

\noindent where \begin{math} \theta \end{math} is the set of parameters of the CNN that the error is calculated from, with \begin{math} \theta_{i} \end{math} being the \begin{math} i \end{math}-th parameter, \begin{math} \phi \end{math} is the weight of the EWC loss, \begin{math} F_{i} \end{math} is the \begin{math} i \end{math}-th diagonal item in the Fisher information matrix, and \begin{math} \theta^{*} \end{math} is the weight set from the previous task used as reference. The matrices were calculated with a single epoch of 262144 image patches. The Fisher information matrices were always calculated using the training set of a single dataset.

The EWC loss was added to the total loss with a weight of \begin{math} \phi = 0.01 \end{math}. In experiments were EWC was used repeatedly, e.g.\ in a multi-task setting, the weight was distributed evenly across the Fisher Information Matrices for each task.

\subsection{Inference and post-processing}
The networks were applied in a fully-convolutional fashion to the WSIs at the level closest to \begin{math}0.5 \times 0.5 \mu m\end{math} pixel spacing \citep{Shelhamer17}. The output of the networks was a whole-slide likelihood map with pixel-wise probabilities (range 0.0–1.0) of being cancerous. Non-tissue areas were masked using the previously obtained tissue masks. 

We used test-time augmentation to improve the robustness of our networks. With this technique the predictions are calculated as an average of multiple augmented versions of the same input. We took the geometric average of the 8 orientations produced by the combinations of vertical mirroring and rotations by multiples of \begin{math} 90^\circ \end{math} for each image. The geometric average has been shown to work slightly better than the arithmetic mean with WSIs \citep{Liu18}.

To translate the likelihood map into detections of individual metastatic lesions, we used non-maxima suppression \citep{Ciresan13} with \begin{math} r = 150 \mu m \end{math} radius on each map. The algorithm was composed of the repetition the following two steps until the global maximum of the probability map fell below 0.5: 1. Find the global maximum \begin{math} m \end{math} and its \begin{math} (x, y) \end{math} coordinates on the probability map and add it to the list of reported points of the WSI. 2. Set the value of the probability map around \begin{math} (x, y) \end{math} with \begin{math} r \end{math} radius to 0.0. This results in a list of coordinates and their corresponding likelihoods.

Furthermore, we also want to generate a slide-level score for one or more metastases being present in the entire slide. We simply picked the highest value from the probability map of each WSI to represent the probability of containing a metastasis for the entire slide. 

For the breast dataset, to be able to calculate the CAMELYON17 challenge metric, Cohen's kappa, we also needed to determine the size of the largest metastasis in each slide. We used a random-forest classifier to classify each slide, using features derived from the cancer probability map, into one of four classes (\textit{negative, ITC, micro,} and \textit{macro}). To derive the features, we thresholded the probability maps at \begin{math} p = 0.5 \end{math} and measured the diameter, area, maximal and mean probability of each resulting region. We used the features derived from the region with the largest diameter and the reference label from each WSI in the 280 slides of the RFC training set to train RFCs. The remaining 120 slides of the RFC validation set were used to evaluate the RFCs. We trained one RFC per CNN in all experiments. The RFCs had 100 tree estimators and the metastasis size class weights were balanced to correct for their uneven ratios.

\subsection{Evaluation metrics}
The overall goal of this paper is to detect metastases in lymph nodes originating from varying cancer types with a single machine learning algorithm. Every case consists of multiple slides, possibly containing several lesions. A comprehensive approach was used, assessing network performance on lesion, slide, and case level. The following subsection details the definition of the used metrics.

\subsubsection{FROC and ROC}
The first two metrics are the exact same as used in the CAMELYON16 challenge and are based on free-response receiver-operating characteristic (FROC) analysis for lesion-level results \citep{Chakraborty11} and receiver-operating characteristic (ROC) analysis for slide level results. This allows us to directly compare our results to those obtained in the challenge.

For FROC analysis, we evaluate each detected point returned by the algorithm with respect to the annotated reference standard. Points outside any annotation are considered false-positives, whereas annotated regions without any corresponding points are considered false negatives. Points within an annotation are true positives. The FROC curve, as defined in the CAMELYON challenge, shows the lesion-level, true-positive fraction (sensitivity) relative to the mean number of false-positive detections in metastasis-free slides. Furthermore, a single score was defined as the average sensitivity across 6 predefined false-positive rates: \begin{math} \frac{1}{4} \end{math} (1 false-positive result in every 4 WSIs), \begin{math} \frac{1}{2} \end{math}, 1, 2, 4, and 8 false-positive findings per WSI.

For the ROC analysis, we calculated the area under the ROC curve (AUC) from the per-WSI probabilities obtained from the algorithms and the per-slide reference labels. Both the FROC and the ROC analysis were also applied to the experiments on the colon and head-and-neck metastasis datasets.

\subsubsection{Cohen's Kappa}
While CAMELYON16 included evaluation on the lesion and slide level, CAMELYON17 was evaluated at the case level. Each case in the challenge was a composition of 5 WSIs, each representing a single lymph node of a patient. Based on the combination of slide-level metastasis labels within a case, a pN stage per case can be determined (Table \ref{table:pn_staging}) \citep{Sobin11}.

\begin{table}
  \centering
  \begin{tabular}{ll}
    \toprule
    \textbf{pN-Stage} & \textbf{Slide Labels}                                  \\
    \midrule
    pN0               & No micro-metastases or macro-metastases or ITC found.  \\
    pN0(i+)           & Only ITC found.                                        \\
    pN1mi             & Micro-metastases found, but no macro-metastases found. \\
    pN1               & Metastases found in 1 -- 3 lymph nodes, of which       \\
                      & at least 1 is a macro-metastasis.                      \\
    pN2               & Metastases found in 4 –- 9 lymph nodes, of which       \\
                      & at least 1 is a macro-metastasis.                      \\
    \bottomrule
  \end{tabular} 
  \caption{pN-stages used in the CAMELYON17 challenge}
  \label{table:pn_staging}
\end{table}

Algorithms participating in CAMELYON17 were compared on their ability to correctly predict the case-level pN stage for each case. As a metric, Cohen's kappa with 5 classes and quadratic weights was used, which measures inter-observer agreement for categorical variables \citep{Fleiss73}. The kappa score was calculated over the pN-stages of the 100 test patients and the output of the algorithms. The kappa metric ranges from \begin{math} -1 \end{math} to \begin{math} +1 \end{math}: a negative value indicates lower than chance agreement, zero indicates exact chance agreement, and a positive value indicates better than chance agreement.

For direct comparison to the algorithms participating in the CAMELYON17 challenge, we convert our RFC predicted metastasis size predictions per CNN to a pN stage per case by applying the rules from Table \ref{table:pn_staging} directly and subsequently calculating the Kappa score.

\subsubsection{Bootstrapping}
For all metrics confidence intervals were obtained using bootstrapping of the test set, with 10000 samples.

\subsection{Experiments}
First, we trained one separate DenseNet on each dataset to obtain the baseline metrics and the cross-domain performance. We refer to these networks as \textit{Specialized Networks}, as they are only taught the appearance of a single cancer type. The Specialized 1, 2 and 3 networks were trained on the breast, colon and head-and-neck datasets, respectively.

To assess potential strategies for `domain adaptation' when training data for the `source task' is available, we perform two experiments, training so-called `Generic' networks and `Extended' networks. We trained two DenseNets, the \textit{Generic Networks} on combined datasets to assess the benefits of larger, more diverse datasets. The Generic 1 network was trained on the union of the breast and colon datasets, while the Generic 2 network was trained on the union of breast, colon and the the head-and-neck datasets. Note that all networks are still optimized for predicting the presence or absence of metastasis, regardless of cancer type.

The \textit{Extended Networks} were trained in multiple stages. In each stage, the training dataset was extended with the addition of a new dataset and the weights of the network of the previous step were used as initialization. The Extended 1 network was initialized with the weights of Specialized 1 network and trained on the combined breast and colon datasets. The Extended 2 includes a third stage in which the combined breast, colon and the the head-and-neck datasets was used for training while the network was initialized using the weights of the Extended 1 network. By using the largest dataset with the highest quality labels for initial pre-training, we expect the network to be in a better local optimum after training compared to the Generic networks, which should result in higher performance across tasks.

Domain adaptation strategies without access to the training data for the `source task' where also investigated. As a baseline we trained, the \textit{Transferred Network} (Transferred 1), which was initialized with the weights of the `Specialized 1 network' and fine-tuned on the colon dataset only. We expect this network to suffer from catastrophic forgetting, i.e.\ losing significant performance on the `source task'.

In our final sets of experiments, we tried to mitigated catastrophic forgetting using EWC. The so-called \textit{Adapted Networks} were also trained in multiple stages, similar to Transferred 1, but with EWC. The Adapted 1 network was first trained on the breast dataset, then adapted to the colon dataset. The Adapted 2 network was trained in three stages. We start from the Adapted 1 network and adapt it to the head-and-neck dataset. All experiments are summarized in Table \ref{table:training_steps}.

After training all networks are evaluated on all test sets using the corresponding metrics:

\begin{itemize}
    \item The CAMELYON16 test set using the FROC and ROC analysis.
    \item The CAMELYON17 test set using the Cohen's kappa analysis.
    \item The colon dataset using the FROC and ROC analysis.
    \item The head-and-neck dataset using the FROC and ROC analysis.
\end{itemize}

\begin{table}
\centering
\setlength{\tabcolsep}{4pt}
\begin{tabular}{l|cccc}
\toprule
\textbf{Network} & \textbf{Step 1} & \textbf{Step 2} & \textbf{Step 3} & \textbf{EWC} \\
\midrule
Specialized 1    & B          & -      & -          & no  \\
Specialized 2    & C          & -      & -          & no  \\
Specialized 3    & HN         & -      & -          & no  \\
\midrule
Generic 1        & B + C      & -      & -          & no  \\
Generic 2        & B + C + HN & -      & -          & no  \\
\midrule
Extended 1       & B          & B + C  & -          & no  \\
Extended 2       & B          & B + C  & B + C + HN & no  \\
\midrule
Transferred 1    & B          & C      & -          & no  \\
\midrule
Adapted 1        & B          & C      & -          & yes \\
Adapted 2        & B          & C      & HN         & yes \\
\bottomrule
\end{tabular}
\caption{Training steps of the CNNs. B: breast, C: colon, HN: head-and-neck dataset.}
\label{table:training_steps}
\end{table}

\section{Results}
\begin{figure}
\centering
  \begin{subfigure}{0.25\textwidth}
    \includegraphics[width=0.94\textwidth]{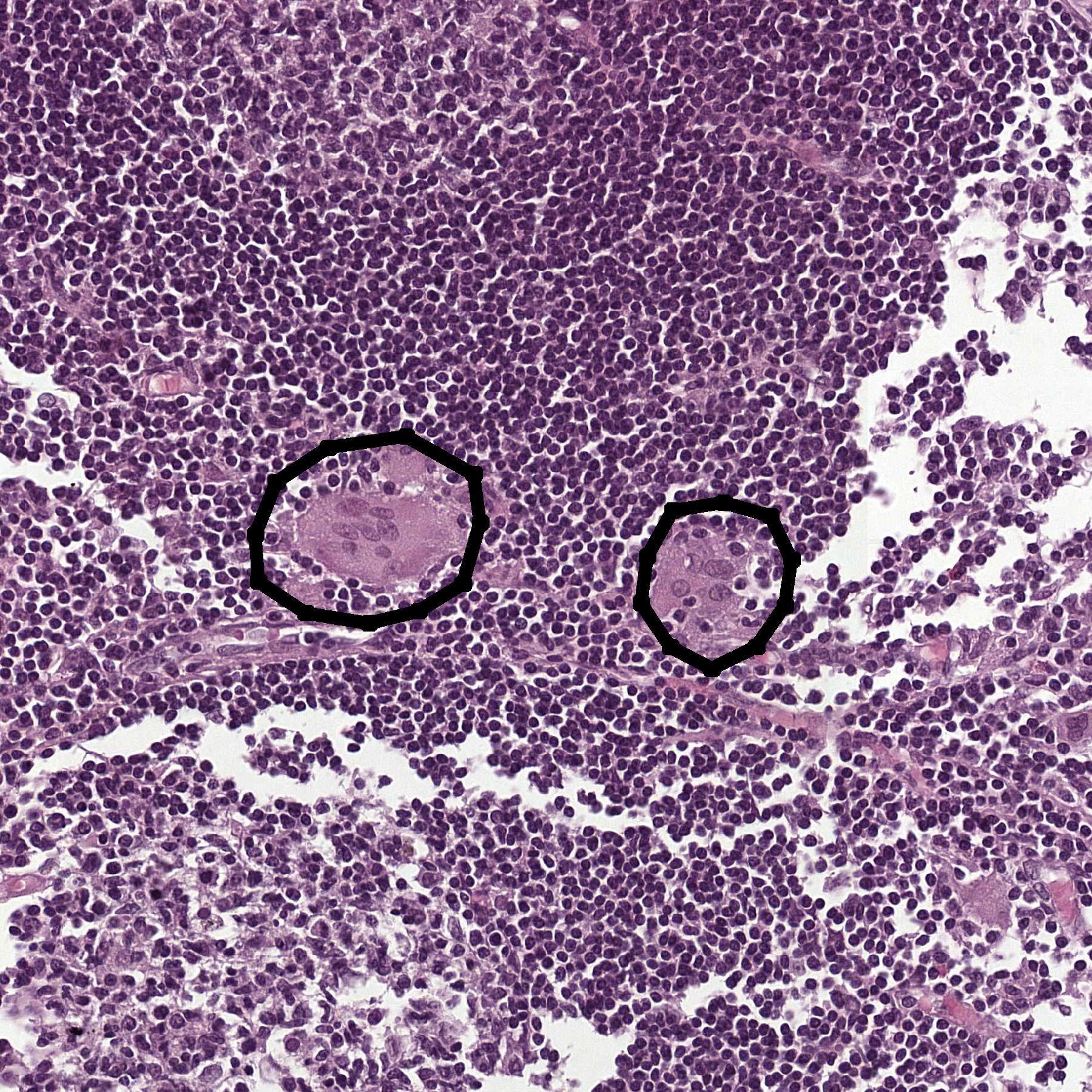}
    \vspace{3.6pt}
  \end{subfigure}%
  \begin{subfigure}{0.25\textwidth}
    \includegraphics[width=0.94\textwidth]{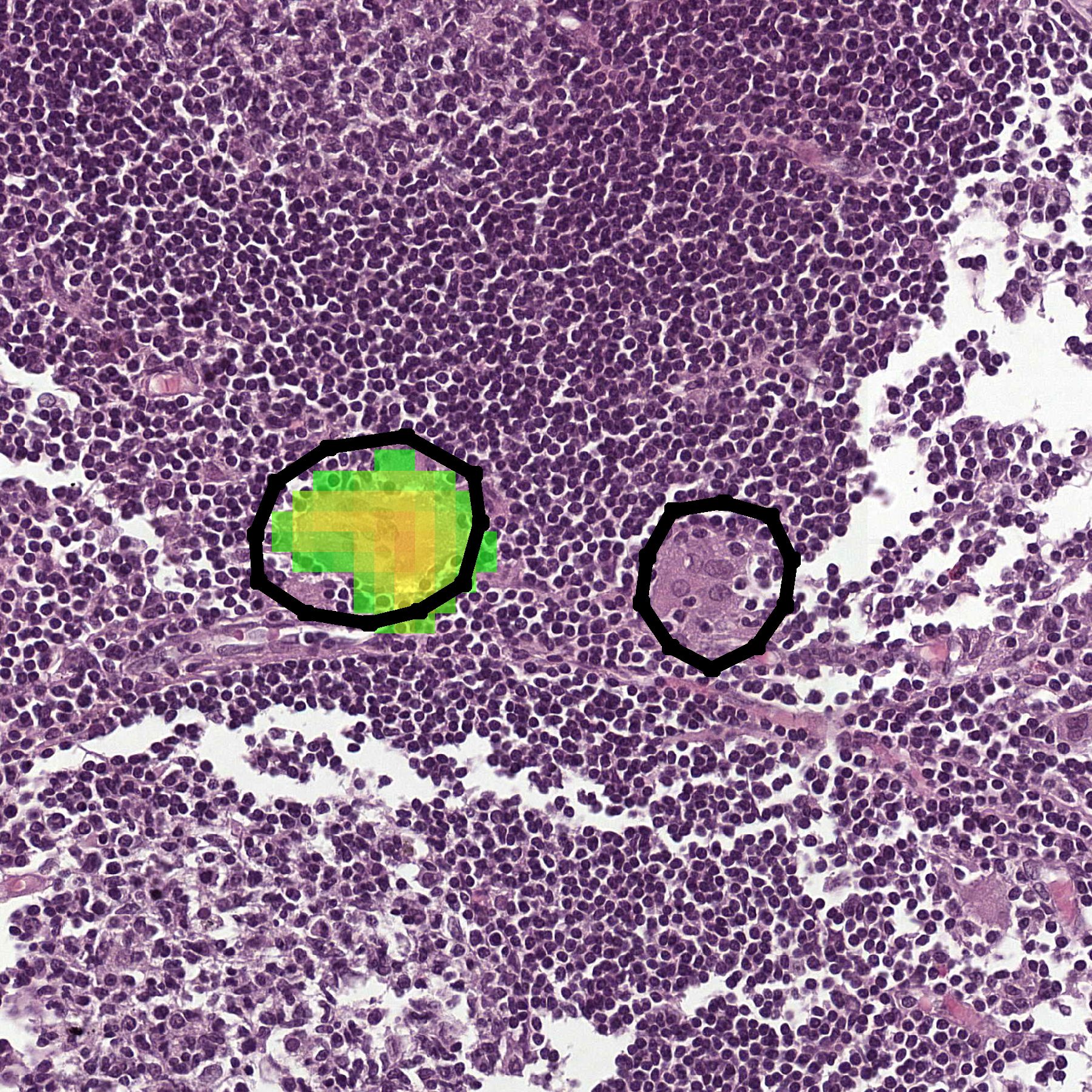}
    \vspace{3.6pt}
  \end{subfigure}%
  \begin{subfigure}{0.25\textwidth}
    \includegraphics[width=0.94\textwidth]{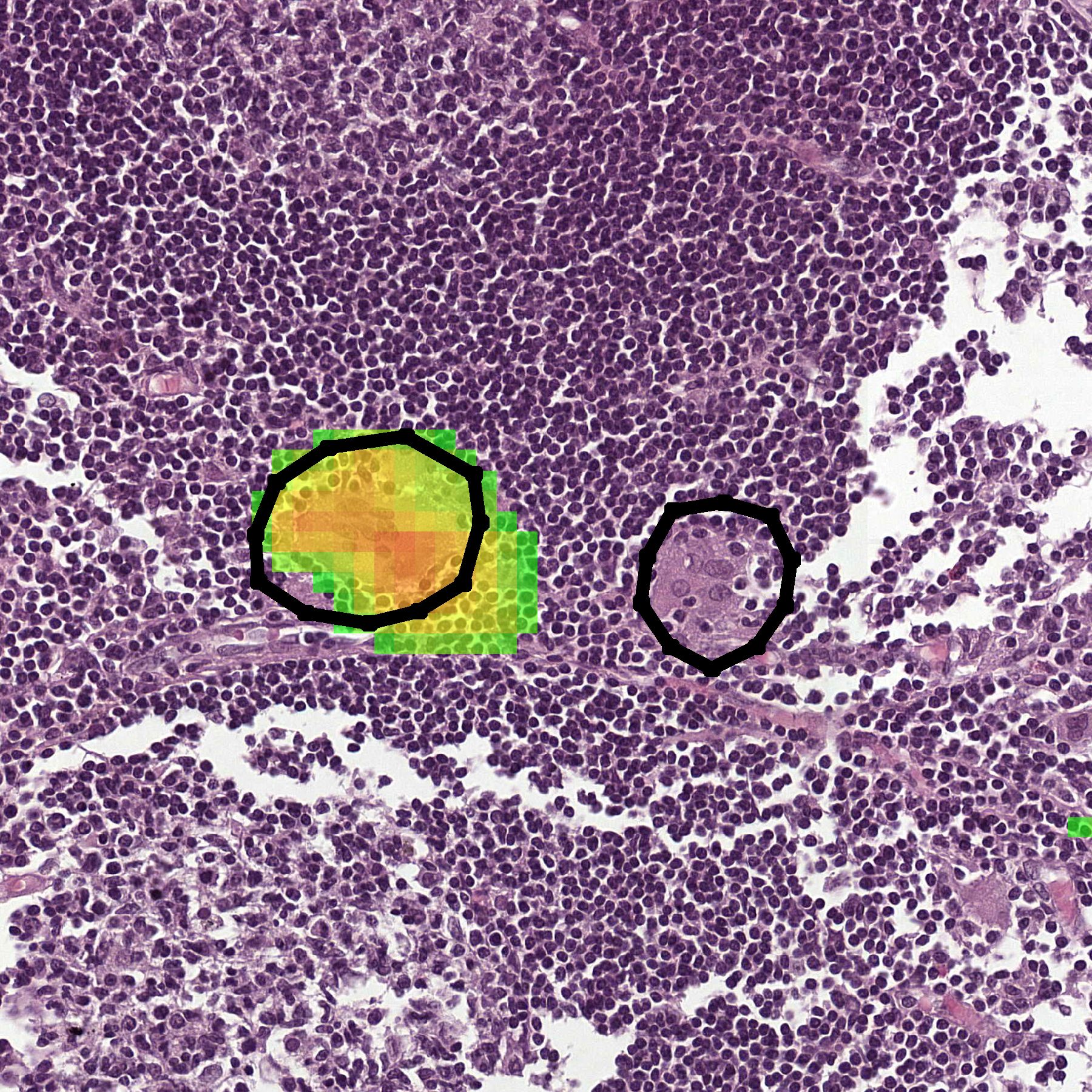}
    \vspace{3.6pt}
  \end{subfigure}%
  \begin{subfigure}{0.25\textwidth}
    \includegraphics[width=0.94\textwidth]{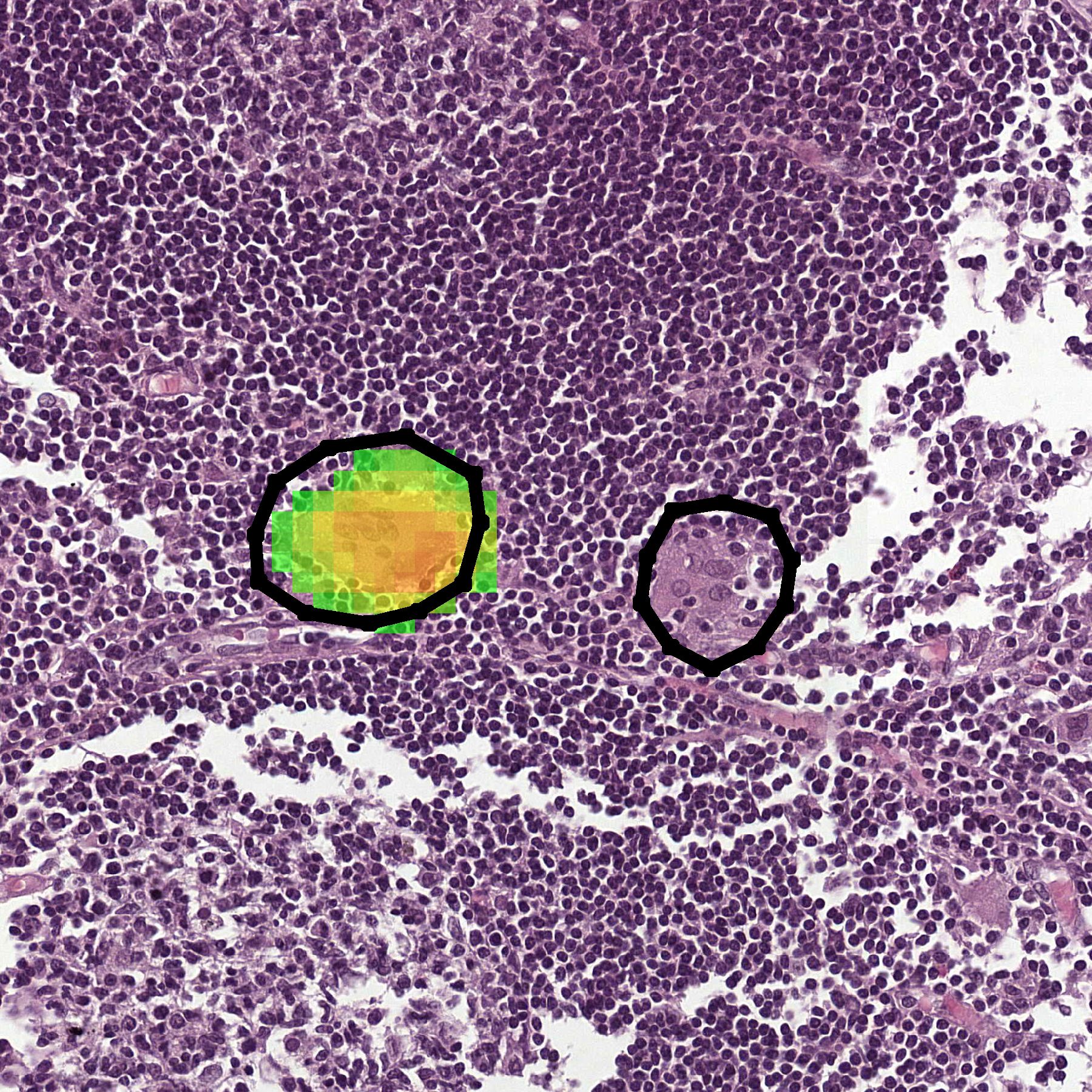}
    \vspace{3.6pt}
  \end{subfigure}\\
  
  \begin{subfigure}{0.25\textwidth}
    \includegraphics[width=0.94\textwidth]{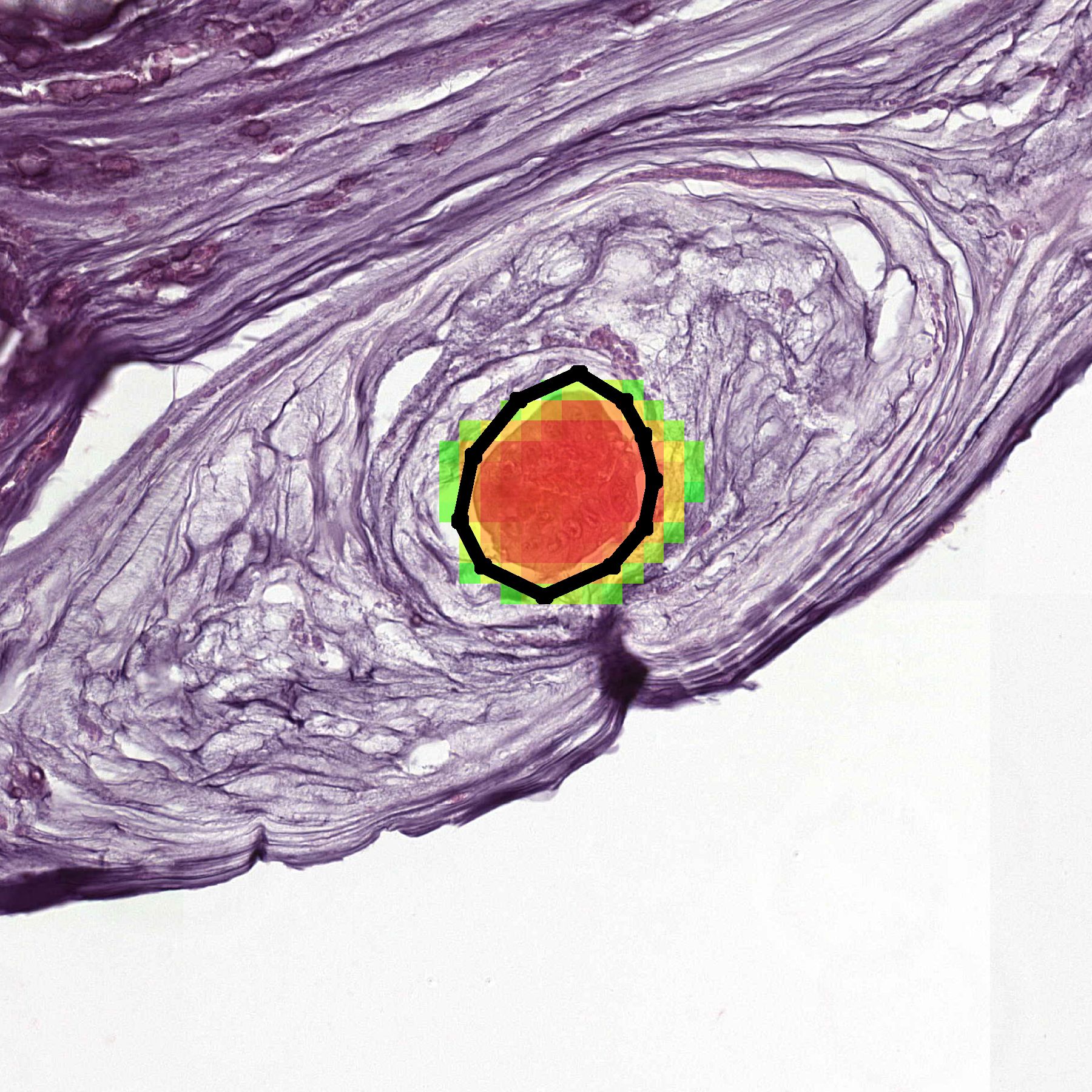}
    \vspace{3.6pt}
  \end{subfigure}%
  \begin{subfigure}{0.25\textwidth}
    \includegraphics[width=0.94\textwidth]{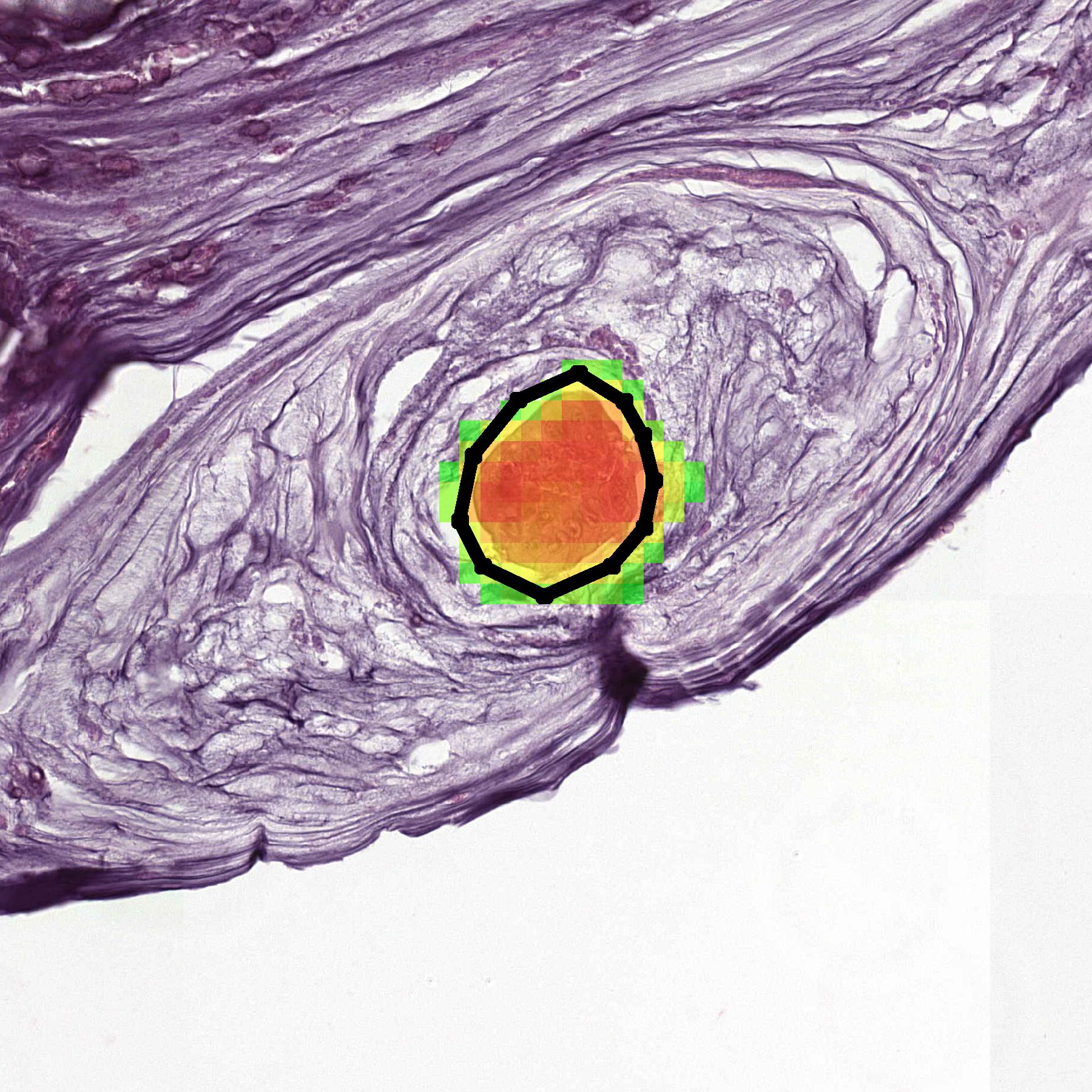}
    \vspace{3.6pt}
  \end{subfigure}%
  \begin{subfigure}{0.25\textwidth}
    \includegraphics[width=0.94\textwidth]{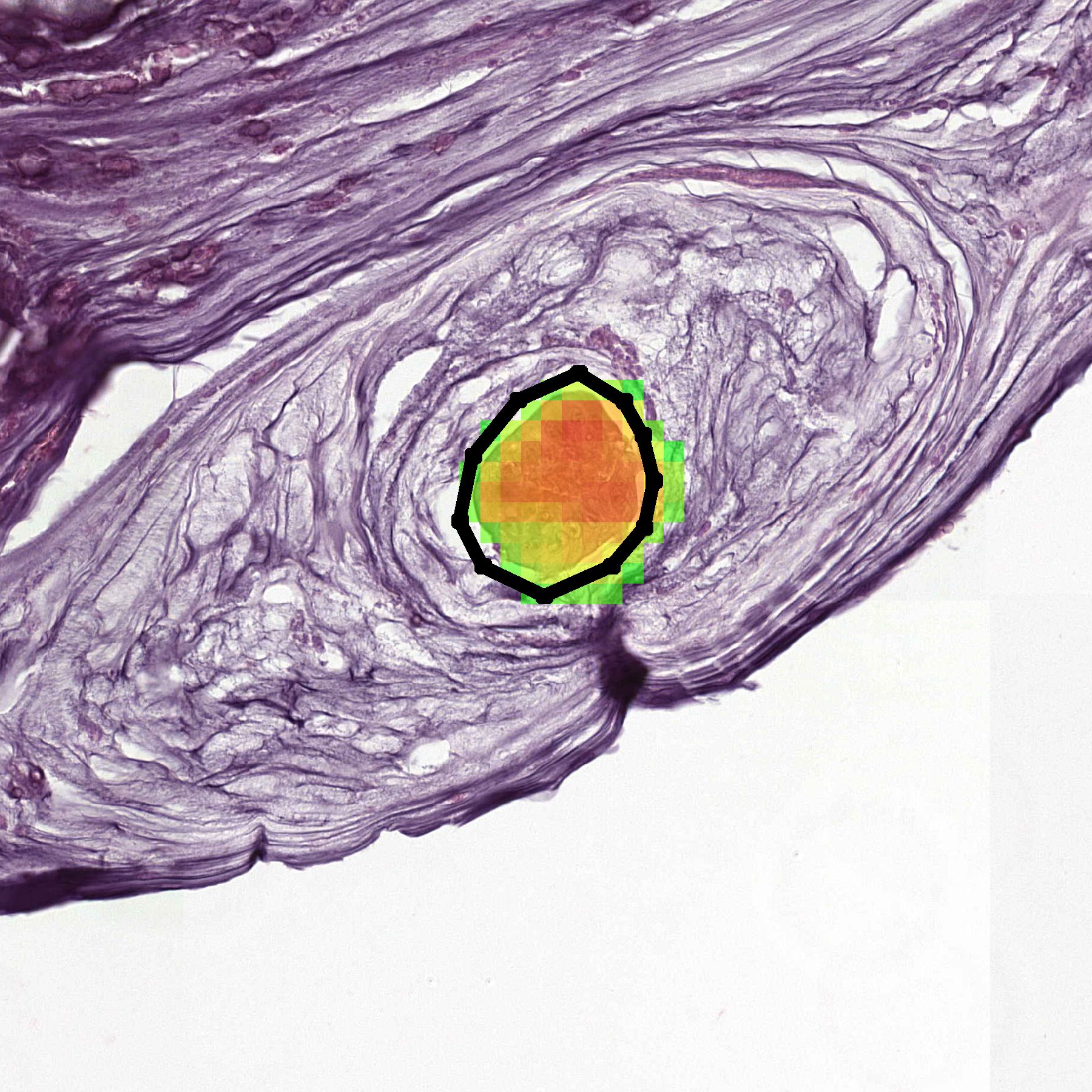}
    \vspace{3.6pt}
  \end{subfigure}%
  \begin{subfigure}{0.25\textwidth}
    \includegraphics[width=0.94\textwidth]{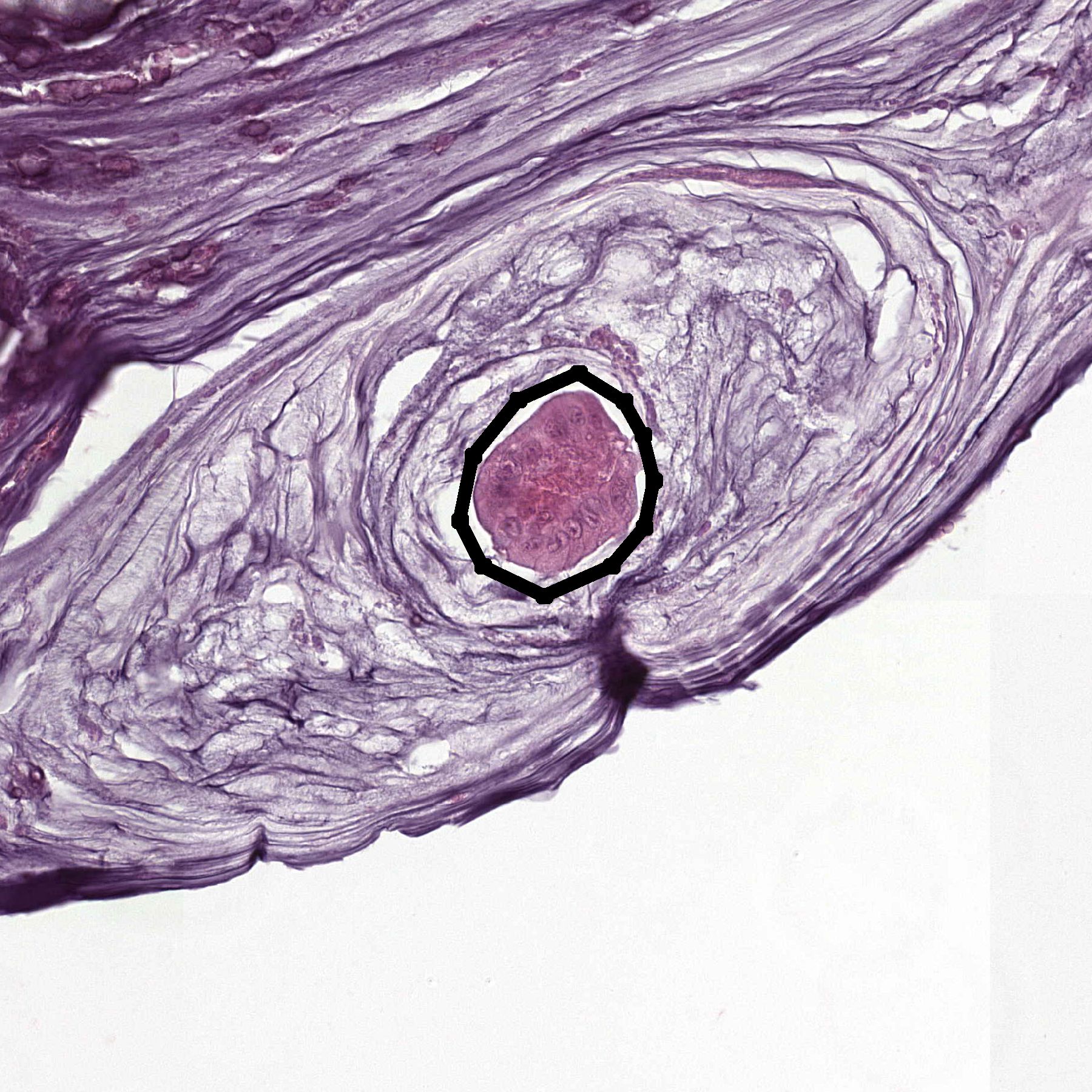}
    \vspace{3.6pt}
  \end{subfigure}\\
  
  \begin{subfigure}{0.25\textwidth}
    \includegraphics[width=0.94\textwidth]{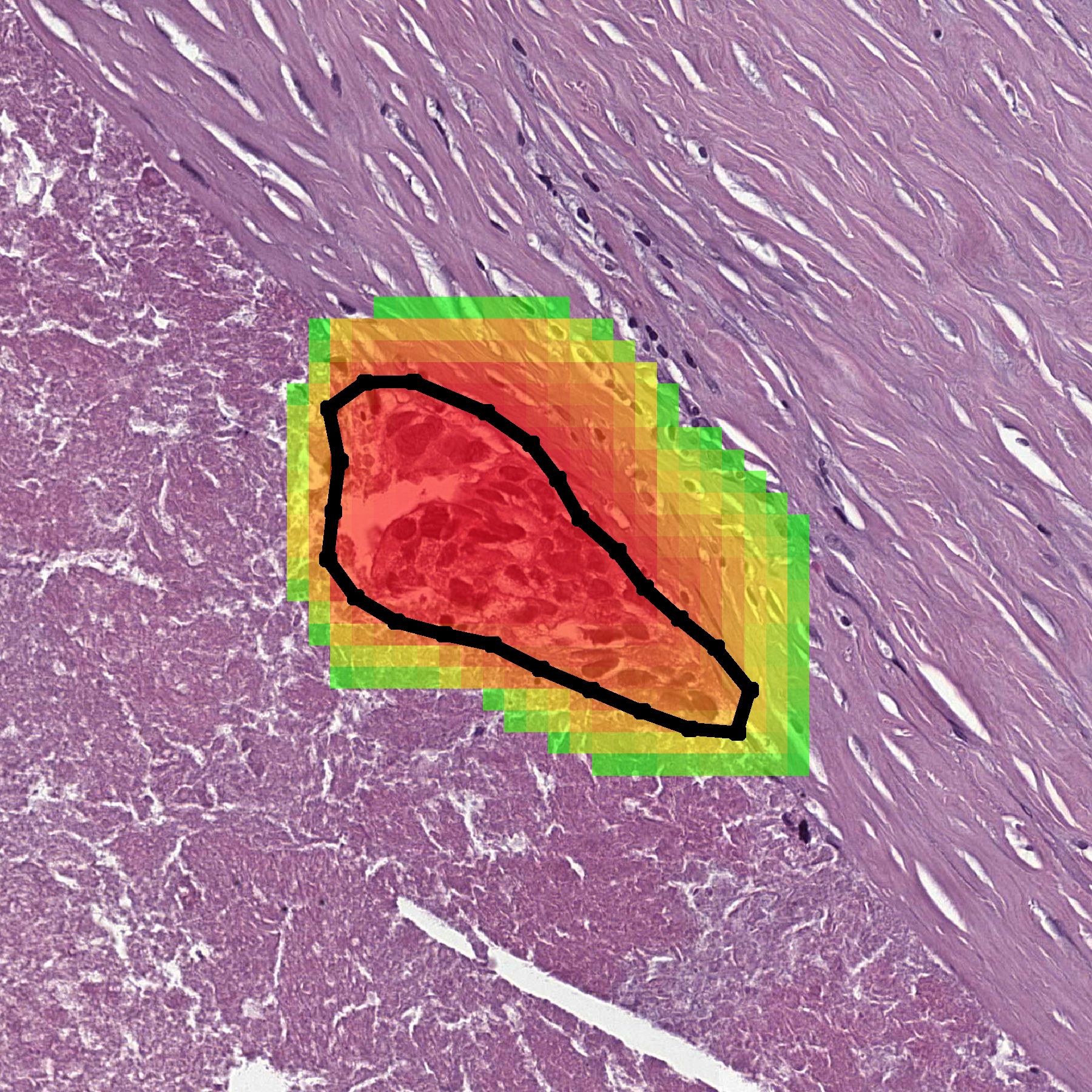}
    \vspace{3.6pt}
  \end{subfigure}%
  \begin{subfigure}{0.25\textwidth}
    \includegraphics[width=0.94\textwidth]{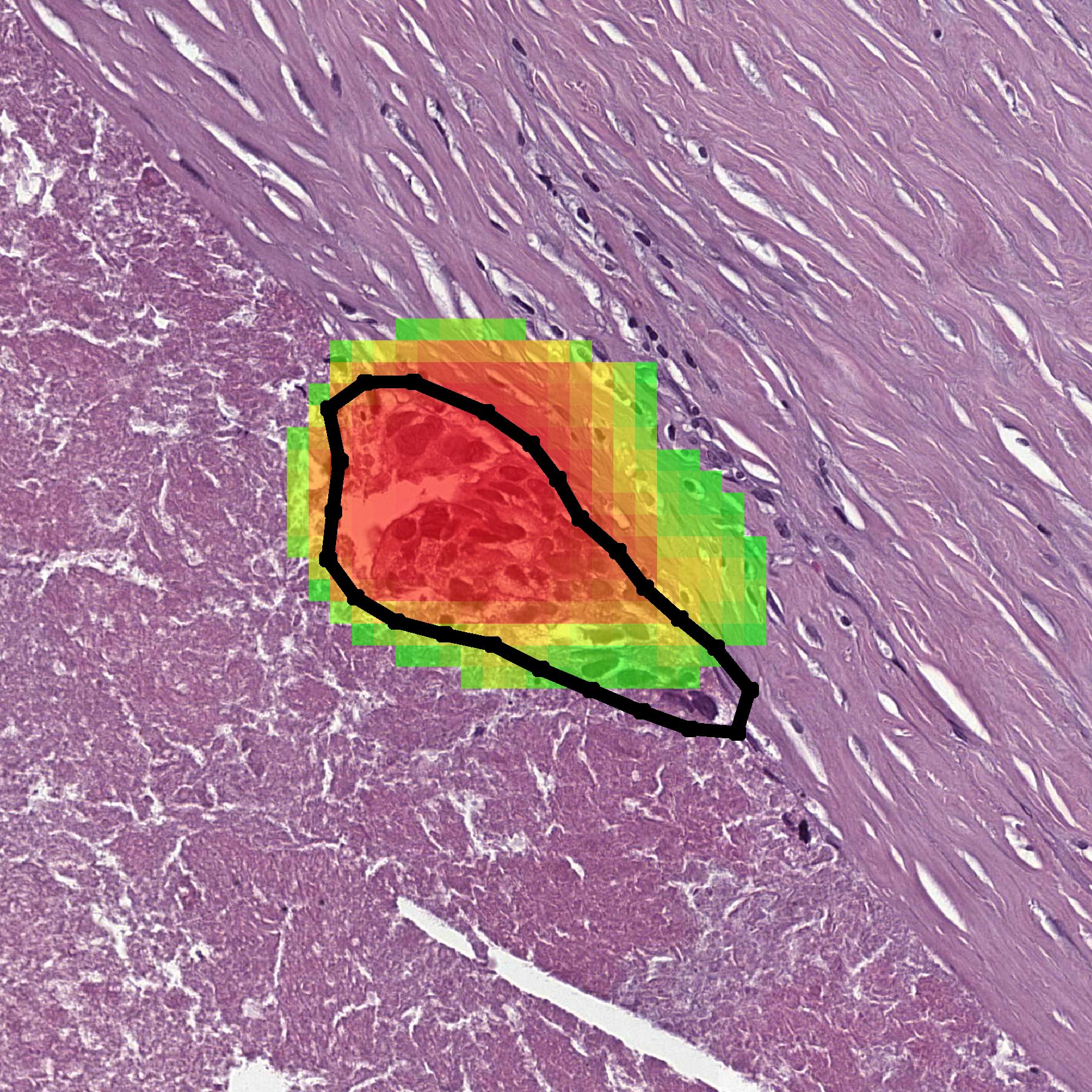}
    \vspace{3.6pt}
  \end{subfigure}%
  \begin{subfigure}{0.25\textwidth}
    \includegraphics[width=0.94\textwidth]{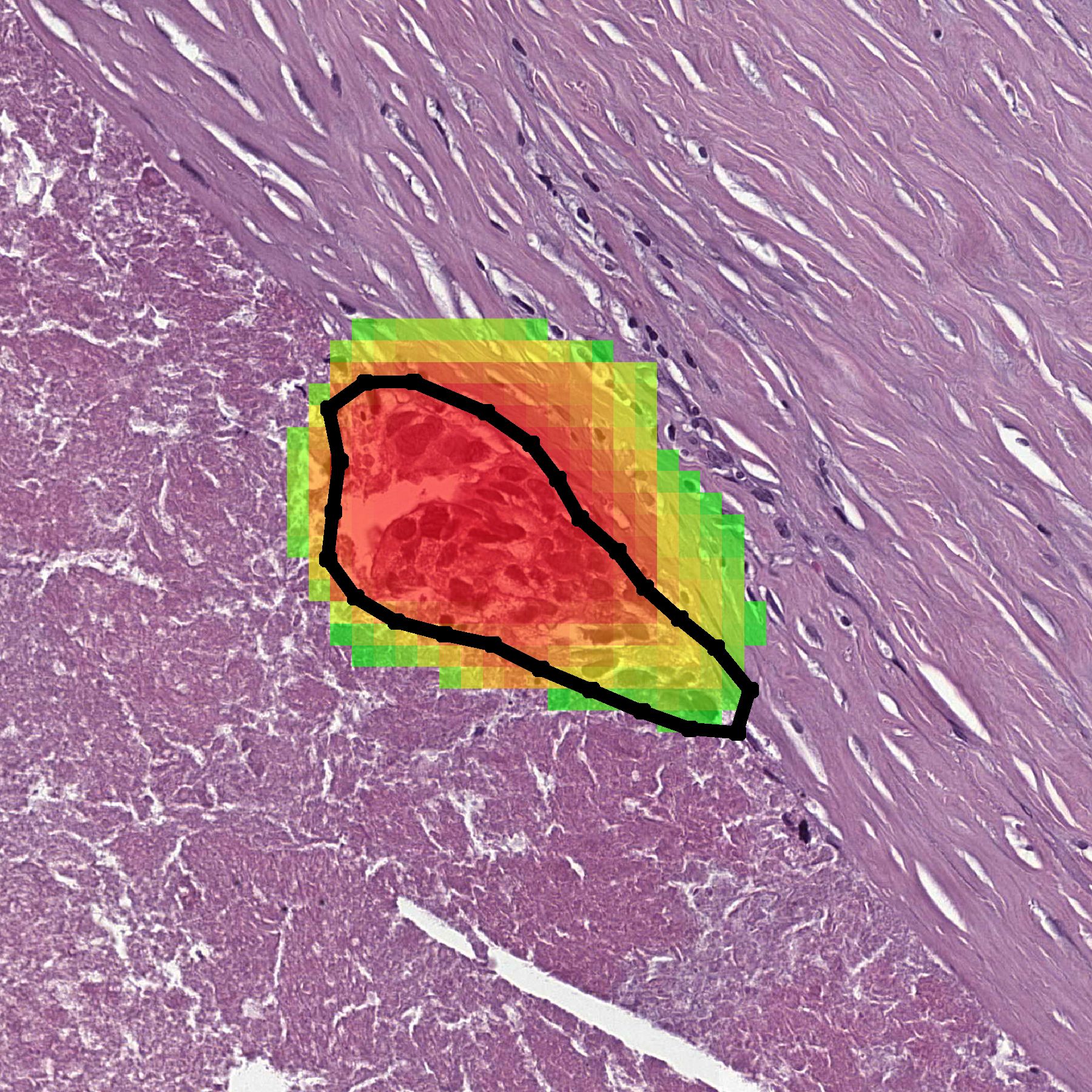}
    \vspace{3.6pt}
  \end{subfigure}%
  \begin{subfigure}{0.25\textwidth}
    \includegraphics[width=0.94\textwidth]{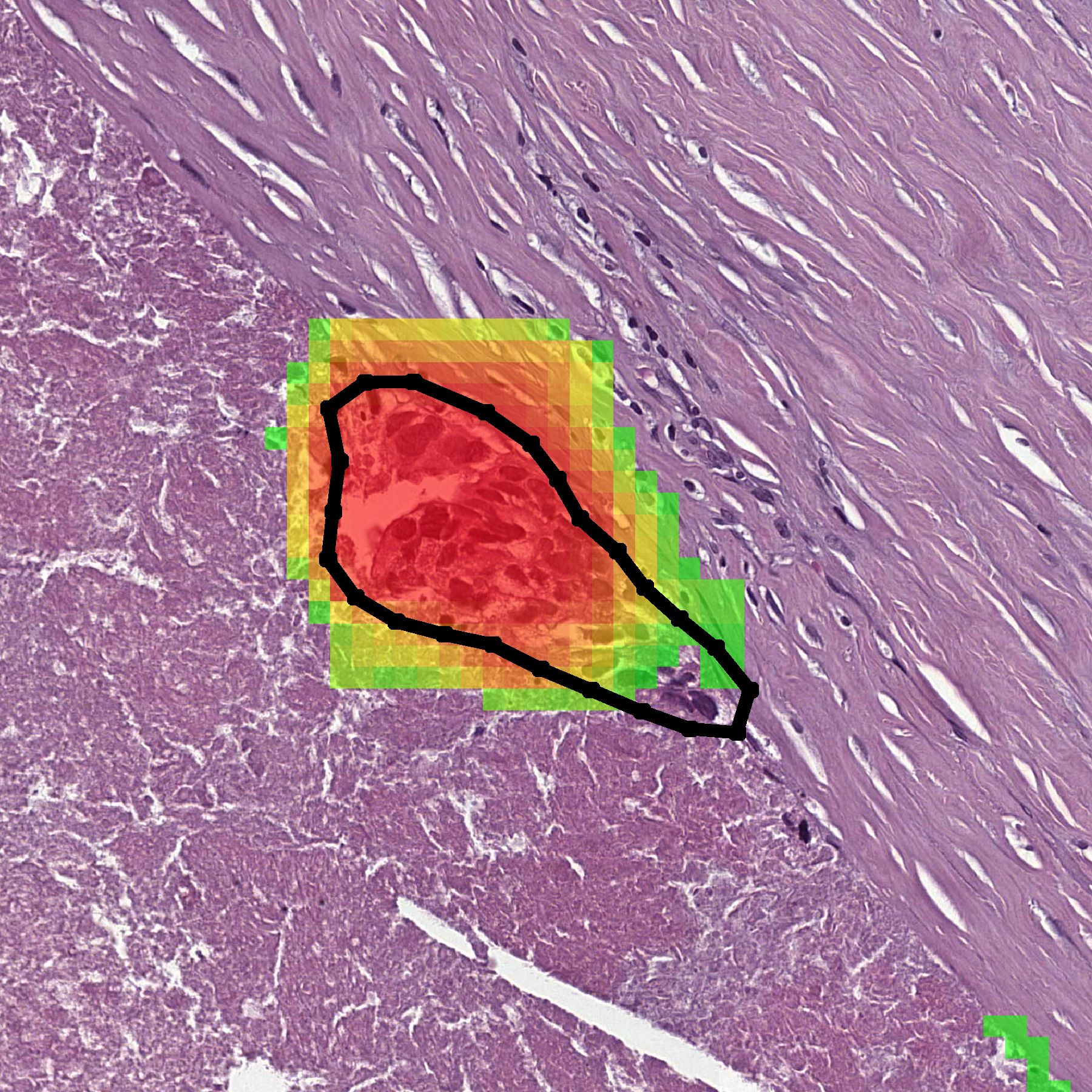}
    \vspace{3.6pt}
  \end{subfigure}\\
  \captionsetup{justification=centering}
  \caption{Examples of likelihood maps. Columns: Specialized 2, Generic 1, Extended 1, and Adapted 1 network outputs. Rows: breast, colon, and head-and-neck lymph node tissue. The colors range from green to red, representing low to high probability respectively. The reference is annotated in black.}
  \label{figure:result_examples}
\end{figure}

All quantitative metric results are presented in Table \ref{table:results}. Examples of likelihood maps are depicted in Figure \ref{figure:result_examples}.

\subsection{Specialized Networks}
For the specialized networks, we can observe mostly expected results: they perform the best on their `source tasks' with Specialized 1 performing the best on the breast dataset and Specialized 2 on the colon dataset with ROC and FROC scores of 0.9690 (ROC), 0.8377 (FROC) and 0.9537 (ROC), 0.7112 (FROC), respectively. The exception is the Specialized 3 network, which does perform best on its `source task' but is outperformed both by Specialized 1 and 2. 

The performance on the other tasks shows more differentiation. Here we can see that the Specialized 1 network, trained on the highest quality source dataset has competitive performance on the colon task (0.02 difference in both ROC and FROC metrics), whereas the Specialized 2 and 3 networks perform poorly in both breast tasks (0.3 - 0.5 drops in FROC and Kappa). 

\subsection{Generic Networks}
The Generic networks are intended to have, on average, better performance across tasks because they were trained with combined datasets. The Generic 1 network confirms this hypothesis, it roughly maintains the ROC, FROC and Kappa scores of the Specialized 1 network on the breast tasks and the best ROC and FROC scores on the head-and-neck task. On the colon task it even outperforms all the Specialized networks. The Generic 2 network show somewhat mixed results with surprising drops in the scores on the breast and head-and-neck tasks.

\subsection{Extended Networks}
For the Extended networks we expected the best performance across tasks as they were pre-trained using the highest quality dataset, which should result in the highest quality local optimum, and then fine-tuned by using the combined datasets. The Extended 1 network had the best performance on the breast and colon tasks on average across the Specialized and Generic networks, with only slightly lower performance (0.02 Kappa) on the CAMELYON17 dataset compared to the Specialized 1 network. Interestingly, adding the head-and-neck dataset in the Extended 2 network again deteriorates performance, similar to the Generic 2 network. The Extended 2 network does outperform the Generic 2 network on all metrics, in line with our hypothesis, except the CAMELYON17 Kappa score, where there is a small drop (0.0054).

\subsection{Transferred and Adapted Networks}

The \textit{Transferred 1} network showed expected results, with catastrophic forgetting causing a large drop in `source task' performance in the CAMELYON16 and 17 metrics with drops of 0.10, 0.33 and 0.33 on the ROC, FROC and Kappa scores, respectively. This shows there is ample room for improvement when fine-tuning networks without the `source task' training data.

For the Adapted networks we see interesting results. As expected, we see that the `source task' performance drop (Specialized 1) is significantly less than without EWC, with drops of only 0.07 and 0.04 in FROC and Kappa scores for Adapted 1 and 0.11 and 0.17 for Adapted 2. Surprisingly the ROC scores even remain roughly equal to Specialized 1. However, Adapted 1 also does not increase performance on the colon task compared to Specialized 1 and 2, but it does increase performance on the head-and-neck dataset (gains of 0.02 and 0.05 in ROC and FROC score). For Adapted 2 we do see improved performance on the colon task (0.04 and 0.06 in ROC and FROC scores) and on the head-and-neck ROC metric (0.06).

\begin{table}
\centering
\tiny
\setlength{\tabcolsep}{3pt}
\begin{tabular}{l|ll|l|ll|ll}
\toprule
\multirow{2}{*}{\textbf{Model}} & \multicolumn{2}{c}{\textbf{CAM. 16}}                              & \multicolumn{1}{c}{\textbf{CAM. 17}} & \multicolumn{2}{c}{\textbf{Colon}}                                   & \multicolumn{2}{c}{\textbf{Head \& Neck}}                            \\
                                & \multicolumn{1}{c}{\textbf{ROC}} & \multicolumn{1}{c}{\textbf{FROC}} & \multicolumn{1}{c}{\textbf{Kappa}}      & \multicolumn{1}{c}{\textbf{ROC}} & \multicolumn{1}{c}{\textbf{FROC}} & \multicolumn{1}{c}{\textbf{ROC}} & \multicolumn{1}{c}{\textbf{FROC}} \\
\midrule
Spec. 1 &  0.9690       &  0.8377       & 0.8394        &  0.9370       &  0.6910       &  0.9089       &  0.8714       \\
        & (0.92 - 1.00) & (0.76 - 0.91) & (0.76 - 0.90) & (0.86 - 0.99) & (0.54 - 0.85) & (0.76 - 1.00) & (0.75 - 0.96) \\
Spec. 2 &  0.8652       &  0.5565       & 0.6522        &  0.9537       &  0.7112       &  0.9475       &  0.8647       \\
        & (0.79 - 0.93) & (0.42 - 0.70) & (0.53 - 0.77) & (0.88 - 1.00) & (0.51 - 0.90) & (0.86 - 1.00) & (0.76 - 0.94) \\
Spec. 3 &  0.8108       &  0.2779       & 0.5185        &  0.8076       &  0.1904       &  0.9342       &  0.7735       \\
        & (0.73 - 0.89) & (0.17 - 0.41) & (0.37 - 0.67) & (0.66 - 0.93) & (0.07 - 0.40) & (0.83 - 1.00) & (0.53 - 0.93) \\
\midrule                                                      
Gen. 1  &  0.9748       &  0.8352       & 0.8287        &  0.9556       &  0.7385       &  0.9053       &  0.8527       \\
        & (0.93 - 1.00) & (0.74 - 0.92) & (0.77 - 0.88) & (0.89 - 0.99) & (0.54 - 0.92) & (0.76 - 1.00) & (0.72 - 0.95) \\
Gen. 2  &  0.9532       &  0.7932       & 0.7529        &  0.9621       &  0.7551       &  0.9579       &  0.8046       \\
        & (0.90 - 1.00) & (0.69 - 0.89) & (0.65 - 0.84) & (0.90 - 1.00) & (0.54 - 0.93) & (0.88 - 1.00) & (0.61 - 0.93) \\
\midrule                                                      
Ext. 1  &  0.9785       &  0.8513       & 0.8143        &  0.9790       &  0.7612       &  0.9157       &  0.8400       \\
        & (0.94 - 1.00) & (0.78 - 0.93) & (0.76 - 0.87) & (0.94 - 1.00) & (0.55 - 0.94) & (0.78 - 1.00) & (0.68 - 0.94) \\
Ext. 2  &  0.9686       &  0.8185       & 0.7475        &  0.9769       &  0.7662       &  0.9584       &  0.8106       \\
        & (0.93 - 1.00) & (0.72 - 0.91) & (0.63 - 0.84) & (0.93 - 1.00) & (0.56 - 0.94) & (0.89 - 1.00) & (0.64 - 0.92) \\
\midrule                                                      
Tr. 1   &  0.8646       &  0.5072       & 0.5075        &  0.9726       &  0.7796       &  0.9160       &  0.7888       \\
        & (0.79 - 0.93) & (0.39 - 0.64) & (0.35 - 0.64) & (0.92 - 1.00) & (0.59 - 0.93) & (0.77 - 1.00) & (0.59 - 0.93) \\
\midrule                                                      
Ad. 1   &  0.9740       &  0.7693       & 0.7926        &  0.9451       &  0.6869       &  0.9267       &  0.9279       \\
        & (0.94 - 1.00) & (0.66 - 0.87) & (0.73 - 0.85) & (0.87 - 0.99) & (0.48 - 0.88) & (0.80 - 1.00) & (0.83 - 0.98) \\
Ad. 2   &  0.9748       &  0.7283       & 0.6771        &  0.9768       &  0.7523       &  0.9687       &  0.9098       \\
        & (0.94 - 1.00) & (0.63 - 0.83) & (0.56 - 0.77) & (0.93 - 1.00) & (0.59 - 0.90) & (0.91 - 1.00) & (0.80 - 0.97) \\
\bottomrule
\end{tabular}
\caption{Metastasis classification and identification results: ROC scores of metastasis classification and FROC scores of metastasis identification both with 95\% confidence intervals. As expected specialized networks are the best on their `source tasks'. The Generic networks have better performance across tasks because they were trained with combined datasets. The Extended networks performed even better than their Generic counterparts showing the advantage of their training scheme. The Transferred network expressed catastrophic forgetting while the Adapted networks shows how it can be mitigated.}
\label{table:results}
\end{table}

\section{Discussion}
In this study we evaluated several strategies for `domain adaptation' for related tasks in histopathology, both when `source task' data is and is not available. When `source task' data is available, leveraging this in a multi-step training scheme can achieve high `target task' performance while also maintaining high `source task' performance. This strategy outperformed training from scratch with the combined dataset and can also be repeated for additional tasks. Our experiments also resulted in state-of-the-art performance on metastasis detection in colon and head-and-neck cancer. Last, we show that catastrophic forgetting can effectively be mitigated using methods such as elastic weight consolidation when `source task' data is not available.

Out of the three datasets the breast dataset had the highest quality. It had three times more images, with more accurately outlined tumor areas, and without the presence of necrosis, mucus, or tissue types other than lymph node or fatty tissue. This is evidenced by the fact that the Specialized 1 network, which was solely trained on this dataset, performs well across tasks and has thus learned a good representation of the `non-metastasis' class. This also shows that, up to a certain level, using a single high-quality dataset that can `teach' the `non-diseased' class can be used to train an `abnormality' detector. The other 2 Specialized networks show expected results, with their best performance on their `source tasks', although the Specialized 3 network is outperformed by both Specialized 1 and 2, showing the limitations of a small dataset for adequately learning the class representations.

The Generic and Extended networks show that when the `source task' training dataset is present, networks can effectively be taught to perform well across `target tasks' without significant loss of performance on the `source task'. Both Generic and Extended 1 maintain high performance on both the breast, colon and head-and-neck tasks, where Extended 1 has the edge in most cases, highlighting the two-stage training strategy as the most effective `domain adaptation' technique when all data is available. An interesting observation is the consistent drop in performance across tasks when adding the head-and-neck dataset in both the Generic and Extended 2 networks. Although we do not have certainty on what causes this effect, we hypothesize that this might be due to the difference in originating cell types of breast and colon cancer compared to head-and-neck cancer. Specifically, breast and colon cancer are generally adenocarcinomas, whereas head-and-neck cancer are typically squamous cell carcinomas. This might make fitting these entities in a single metastatic class more difficult. In future work we will investigate whether taking into account the originating cell types in `domain adaptation' for histopathology could help boost performance.

Another important observation resulting from the Generic and Extended experiments is that, if we have only a small dataset for a `target task', including a large, high-quality dataset in the training process is an effective way to boost performance, as can be seen by comparing the results of the Specialized 2 and 3 networks to Generic and Extended 1 on the colon and head-and-neck tasks. This strategy is even competitive to fine-tuning a pre-trained network, which suffers from catastrophic forgetting, as can be seen by comparing the colon task results of Extended 2 to Transferred 1 (0.004 and -0.013 differences in ROC and FROC scores). 

Our second set of experiments focused on mitigating catastrophic forgetting in related tasks when the `source task' data is not available, or when re-training the network is not feasible. As shown by the Transferred 1 results, naively fine-tuning a pre-trained network destroys performance on the `source task', but the Adapted 1 and 2 results show that this can effectively be mitigated, with more than 90\% of the performance maintained for the Adapted 1 network and 80\% for Adapted 2, which was fine-tuned twice on new tasks. A surprising result was the fact that the Adapted 2 network showed higher performance on not just the head-and-neck, but also the colon task. We hypothesize this can be caused by the fact that in Adapted 1, the appearance of metastases across the task is more similar. With the added loss on weight changes due to EWC, the weights might not change value significantly during fine-tuning. For Adapted 2, which includes fine-tuning on the head-and-neck dataset, which is more dissimilar, there is probably more pressure to alter the weight values, allowing them to deviate more from their source task values, resulting more generic features which perform better across tasks. This is supported by the larger drop in `source task' performance compared to Adapted 1 and the gains in the `target tasks'.

Overall, our study shows that existing high-quality histopathological datasets can be leveraged for closely related tasks. It demonstrates that a single network can be effectively trained to detect multiple types of abnormalities in a single type of tissue and that it is possible to solve tasks with limited amount of data with a help of a large related dataset.

\section*{Funding}
This work was funded by the Automation in Medical Imaging project. Funding sources were Radboud University Medical Center and Fraunhofer-Gesellschaft. The funders had no role in study design, data collection and analysis, decision to publish, or preparation of the manuscript.

\bibliographystyle{model2-names.bst}
\biboptions{authoryear}
\bibliography{references}

\end{document}